\documentstyle[12pt]{article}

\expandafter\ifx\csname amssym.def\endcsname\relax \else\endinput\fi
\expandafter\edef\csname amssym.def\endcsname{%
       \catcode`\noexpand\@=\the\catcode`\@\space}
\catcode`\@=11
\def\undefine#1{\let#1\undefined}
\def\newsymbol#1#2#3#4#5{\let\next@\relax
 \ifnum#2=\@ne\let\next@\msafam@\else
 \ifnum#2=\tw@\let\next@\msbfam@\fi\fi
 \mathchardef#1="#3\next@#4#5}
\def\mathhexbox@#1#2#3{\relax
 \ifmmode\mathpalette{}{\m@th\mathchar"#1#2#3}%
 \else\leavevmode\hbox{$\m@th\mathchar"#1#2#3$}\fi}
\def\hexnumber@#1{\ifcase#1 0\or 1\or 2\or 3\or 4\or 5\or 6\or 7\or 8\or
 9\or A\or B\or C\or D\or E\or F\fi}

\ifcase\@ptsize
  \font\tenmsa=msam10
  \font\sevenmsa=msam7
  \font\fivemsa=msam5
\or
  \font\tenmsa=msam10  scaled \magstephalf
  \font\sevenmsa=msam7 scaled \magstephalf
  \font\fivemsa=msam5  scaled \magstephalf
\or
  \font\tenmsa=msam10  scaled \magstep1
  \font\sevenmsa=msam7 scaled \magstep1
  \font\fivemsa=msam5  scaled \magstep1
\fi

\newfam\msafam
\textfont\msafam=\tenmsa
\scriptfont\msafam=\sevenmsa
\scriptscriptfont\msafam=\fivemsa
\edef\msafam@{\hexnumber@\msafam}
\mathchardef\dabar@"0\msafam@39
\def\dashrightarrow{\mathrel{\dabar@\dabar@\mathchar"0\msafam@4B}}
\def\dashleftarrow{\mathrel{\mathchar"0\msafam@4C\dabar@\dabar@}}

\def\ulcorner{\delimiter"4\msafam@70\msafam@70 }
\def\urcorner{\delimiter"5\msafam@71\msafam@71 }
\def\llcorner{\delimiter"4\msafam@78\msafam@78 }
\def\lrcorner{\delimiter"5\msafam@79\msafam@79 }
\def\yen{{\mathhexbox@\msafam@55 }}
\def\checkmark{{\mathhexbox@\msafam@58 }}
\def\circledR{{\mathhexbox@\msafam@72 }}
\def\maltese{{\mathhexbox@\msafam@7A }}

\ifcase\@ptsize
  \font\tenmsb=msbm10
  \font\sevenmsb=msbm7
  \font\fivemsb=msbm5
\or
  \font\tenmsb=msbm10  scaled \magstephalf
  \font\sevenmsb=msbm7 scaled \magstephalf
  \font\fivemsb=msbm5  scaled \magstephalf
\or
  \font\tenmsb=msbm10  scaled \magstep1
  \font\sevenmsb=msbm7 scaled \magstep1
  \font\fivemsb=msbm5  scaled \magstep1
\fi

\newfam\msbfam
\textfont\msbfam=\tenmsb
\scriptfont\msbfam=\sevenmsb
\scriptscriptfont\msbfam=\fivemsb
\edef\msbfam@{\hexnumber@\msbfam}
\def\Bbb#1{{\fam\msbfam\relax#1}}
\def\widehat#1{\setbox\z@\hbox{$\m@th#1$}%
 \ifdim\wd\z@>\tw@ em\mathaccent"0\msbfam@5B{#1}%
 \else\mathaccent"0362{#1}\fi}
\def\widetilde#1{\setbox\z@\hbox{$\m@th#1$}%
 \ifdim\wd\z@>\tw@ em\mathaccent"0\msbfam@5D{#1}%
 \else\mathaccent"0365{#1}\fi}

\ifcase\@ptsize
  \font\teneufm=eufm10
  \font\seveneufm=eufm7
  \font\fiveeufm=eufm5
\or
  \font\teneufm=eufm10   scaled \magstephalf
  \font\seveneufm=eufm7  scaled \magstephalf
  \font\fiveeufm=eufm5   scaled \magstephalf
\or
  \font\teneufm=eufm10   scaled \magstep1
  \font\seveneufm=eufm7  scaled \magstep1
  \font\fiveeufm=eufm5   scaled \magstep1
\fi

\newfam\eufmfam
\textfont\eufmfam=\teneufm
\scriptfont\eufmfam=\seveneufm
\scriptscriptfont\eufmfam=\fiveeufm
\def\frak#1{{\fam\eufmfam\relax#1}}

\csname amssym.def\endcsname

\expandafter\ifx\csname pre amssym.tex at\endcsname\relax \else \endinput\fi
\expandafter\chardef\csname pre amssym.tex at\endcsname=\the\catcode`\@
\catcode`\@=11
\newsymbol\boxdot 1200
\newsymbol\boxplus 1201
\newsymbol\boxtimes 1202
\newsymbol\square 1003
\newsymbol\blacksquare 1004
\newsymbol\centerdot 1205
\newsymbol\lozenge 1006
\newsymbol\blacklozenge 1007
\newsymbol\circlearrowright 1308
\newsymbol\circlearrowleft 1309
\undefine\rightleftharpoons
\newsymbol\rightleftharpoons 130A
\newsymbol\leftrightharpoons 130B
\newsymbol\boxminus 120C
\newsymbol\Vdash 130D
\newsymbol\Vvdash 130E
\newsymbol\vDash 130F
\newsymbol\twoheadrightarrow 1310
\newsymbol\twoheadleftarrow 1311
\newsymbol\leftleftarrows 1312
\newsymbol\rightrightarrows 1313
\newsymbol\upuparrows 1314
\newsymbol\downdownarrows 1315
\newsymbol\upharpoonright 1316
 
\newsymbol\downharpoonright 1317
\newsymbol\upharpoonleft 1318
\newsymbol\downharpoonleft 1319
\newsymbol\rightarrowtail 131A
\newsymbol\leftarrowtail 131B
\newsymbol\leftrightarrows 131C
\newsymbol\rightleftarrows 131D
\newsymbol\Lsh 131E
\newsymbol\Rsh 131F
\newsymbol\rightsquigarrow 1320
\newsymbol\leftrightsquigarrow 1321
\newsymbol\looparrowleft 1322
\newsymbol\looparrowright 1323
\newsymbol\circeq 1324
\newsymbol\succsim 1325
\newsymbol\gtrsim 1326
\newsymbol\gtrapprox 1327
\newsymbol\multimap 1328
\newsymbol\therefore 1329
\newsymbol\because 132A
\newsymbol\doteqdot 132B
 
\newsymbol\triangleq 132C
\newsymbol\precsim 132D
\newsymbol\lesssim 132E
\newsymbol\lessapprox 132F
\newsymbol\eqslantless 1330
\newsymbol\eqslantgtr 1331
\newsymbol\curlyeqprec 1332
\newsymbol\curlyeqsucc 1333
\newsymbol\preccurlyeq 1334
\newsymbol\leqq 1335
\newsymbol\leqslant 1336
\newsymbol\lessgtr 1337
\newsymbol\backprime 1038
\newsymbol\risingdotseq 133A
\newsymbol\fallingdotseq 133B
\newsymbol\succcurlyeq 133C
\newsymbol\geqq 133D
\newsymbol\geqslant 133E
\newsymbol\gtrless 133F
\newsymbol\sqsubset 1340
\newsymbol\sqsupset 1341
\newsymbol\vartriangleright 1342
\newsymbol\vartriangleleft 1343
\newsymbol\trianglerighteq 1344
\newsymbol\trianglelefteq 1345
\newsymbol\bigstar 1046
\newsymbol\between 1347
\newsymbol\blacktriangledown 1048
\newsymbol\blacktriangleright 1349
\newsymbol\blacktriangleleft 134A
\newsymbol\vartriangle 134D
\newsymbol\blacktriangle 104E
\newsymbol\triangledown 104F
\newsymbol\eqcirc 1350
\newsymbol\lesseqgtr 1351
\newsymbol\gtreqless 1352
\newsymbol\lesseqqgtr 1353
\newsymbol\gtreqqless 1354
\newsymbol\Rrightarrow 1356
\newsymbol\Lleftarrow 1357
\newsymbol\veebar 1259
\newsymbol\barwedge 125A
\newsymbol\doublebarwedge 125B
\undefine\angle
\newsymbol\angle 105C
\newsymbol\measuredangle 105D
\newsymbol\sphericalangle 105E
\newsymbol\varpropto 135F
\newsymbol\smallsmile 1360
\newsymbol\smallfrown 1361
\newsymbol\Subset 1362
\newsymbol\Supset 1363
\newsymbol\Cup 1264
 
\newsymbol\Cap 1265
 
\newsymbol\curlywedge 1266
\newsymbol\curlyvee 1267
\newsymbol\leftthreetimes 1268
\newsymbol\rightthreetimes 1269
\newsymbol\subseteqq 136A
\newsymbol\supseteqq 136B
\newsymbol\bumpeq 136C
\newsymbol\Bumpeq 136D
\newsymbol\lll 136E
 
\newsymbol\ggg 136F
 
\newsymbol\circledS 1073
\newsymbol\pitchfork 1374
\newsymbol\dotplus 1275
\newsymbol\backsim 1376
\newsymbol\backsimeq 1377
\newsymbol\complement 107B
\newsymbol\intercal 127C
\newsymbol\circledcirc 127D
\newsymbol\circledast 127E
\newsymbol\circleddash 127F
\newsymbol\lvertneqq 2300
\newsymbol\gvertneqq 2301
\newsymbol\nleq 2302
\newsymbol\ngeq 2303
\newsymbol\nless 2304
\newsymbol\ngtr 2305
\newsymbol\nprec 2306
\newsymbol\nsucc 2307
\newsymbol\lneqq 2308
\newsymbol\gneqq 2309
\newsymbol\nleqslant 230A
\newsymbol\ngeqslant 230B
\newsymbol\lneq 230C
\newsymbol\gneq 230D
\newsymbol\npreceq 230E
\newsymbol\nsucceq 230F
\newsymbol\precnsim 2310
\newsymbol\succnsim 2311
\newsymbol\lnsim 2312
\newsymbol\gnsim 2313
\newsymbol\nleqq 2314
\newsymbol\ngeqq 2315
\newsymbol\precneqq 2316
\newsymbol\succneqq 2317
\newsymbol\precnapprox 2318
\newsymbol\succnapprox 2319
\newsymbol\lnapprox 231A
\newsymbol\gnapprox 231B
\newsymbol\nsim 231C
\newsymbol\ncong 231D
\newsymbol\diagup 231E
\newsymbol\diagdown 231F
\newsymbol\varsubsetneq 2320
\newsymbol\varsupsetneq 2321
\newsymbol\nsubseteqq 2322
\newsymbol\nsupseteqq 2323
\newsymbol\subsetneqq 2324
\newsymbol\supsetneqq 2325
\newsymbol\varsubsetneqq 2326
\newsymbol\varsupsetneqq 2327
\newsymbol\subsetneq 2328
\newsymbol\supsetneq 2329
\newsymbol\nsubseteq 232A
\newsymbol\nsupseteq 232B
\newsymbol\nparallel 232C
\newsymbol\nmid 232D
\newsymbol\nshortmid 232E
\newsymbol\nshortparallel 232F
\newsymbol\nvdash 2330
\newsymbol\nVdash 2331
\newsymbol\nvDash 2332
\newsymbol\nVDash 2333
\newsymbol\ntrianglerighteq 2334
\newsymbol\ntrianglelefteq 2335
\newsymbol\ntriangleleft 2336
\newsymbol\ntriangleright 2337
\newsymbol\nleftarrow 2338
\newsymbol\nrightarrow 2339
\newsymbol\nLeftarrow 233A
\newsymbol\nRightarrow 233B
\newsymbol\nLeftrightarrow 233C
\newsymbol\nleftrightarrow 233D
\newsymbol\divideontimes 223E
\newsymbol\varnothing 203F
\newsymbol\nexists 2040
\newsymbol\Finv 2060
\newsymbol\Game 2061
\newsymbol\mho 2066
\newsymbol\eth 2067
\newsymbol\eqsim 2368
\newsymbol\beth 2069
\newsymbol\gimel 206A
\newsymbol\daleth 206B
\newsymbol\lessdot 236C
\newsymbol\gtrdot 236D
\newsymbol\ltimes 226E
\newsymbol\rtimes 226F
\newsymbol\shortmid 2370
\newsymbol\shortparallel 2371
\newsymbol\smallsetminus 2272
\newsymbol\thicksim 2373
\newsymbol\thickapprox 2374
\newsymbol\approxeq 2375
\newsymbol\succapprox 2376
\newsymbol\precapprox 2377
\newsymbol\curvearrowleft 2378
\newsymbol\curvearrowright 2379
\newsymbol\digamma 207A
\newsymbol\varkappa 207B
\newsymbol\Bbbk 207C
\newsymbol\hslash 207D
\undefine\hbar
\newsymbol\hbar 207E
\newsymbol\backepsilon 237F
\catcode`\@=\csname pre amssym.tex at\endcsname

\def\Box{\hbox{\vrule height1ex\kern-0.4pt
\vbox to 1ex{\hrule width1ex\vfil\hrule width1ex}\kern-0.4pt\vrule height1ex}}
\newcommand{\sqr}[2]{{{\vcenter{\vbox{\hrule height.#2pt
\hbox{\vrule width.#2pt height#1pt \kern#1pt
\vrule width.#2pt}
\hrule height.#2pt}}}}}
\newcommand{\uph}{\upharpoonright}

\newtheorem{prop}{Proposition}
\newtheorem{theorem}{Theorem}
\newcommand{\ovl}{\overline}
\newcommand{\til}{\tilde}

\newcommand{\eo}{\setcounter{equation}{0}}

\newcommand{\be}{\begin{equation}}

\newcommand{\ee}{\end{equation}}
\newcommand{\al}{\alpha}
\newcommand{\bt}{\beta}

\newcommand{\dl}{\delta}
\newcommand{\ep}{\epsilon}

\newcommand{\et}{\eta}
\newcommand{\th}{\theta}
\newcommand{\vth}{\vartheta}

\newcommand{\lm}{\lambda}
\newcommand{\Lm}{\Lambda}
\newcommand{\rh}{\rho}
\newcommand{\sg}{\sigma}
\newcommand{\ta}{\tau}

\newcommand{\phv}{\varphi}

\newcommand{\ps}{\psi}

\newcommand{\om}{\omega}
\newcommand{\Om}{\Omega}
\newcommand{\nn}{\nonumber}
\newcommand{\raw}{\rightarrow}
\newcommand{\A}{\frak A}
\newcommand{\B}{\frak B}
\newcommand{\F}{\frak F}
\newcommand{\C}{{\Bbb C}}

\newcommand{\bib}{\bibitem}
\newcommand{\cin}{C^{\infty}}

\renewcommand{\H}{\mbox{$\cal H$}}

\newcommand{\n}{\parallel}

 \renewcommand{\ll}{\label}

\newcommand{\R}{{\Bbb R}}

\newcommand{\bop}{{\mbox{{\bf p}}}}

\newcommand{\notp}{p \kern-.48em /}
\newcommand{\ci}{\cite}
\newcommand{\bea}{\begin{eqnarray}}
\newcommand{\eea}{\end{eqnarray}}
\newcommand{\ot}{\otimes}
\newcommand{\half}{\mbox{\footnotesize $\frac{1}{2}$}}
\topmargin = - 0.5 cm
\newcommand{\rip}{(\cdot,\cdot)_0}
\newcommand{\pco}{\al_{\rm co}}
\newcommand{\bolt}{{\bf l}_{\tilde{p}}}
\newcommand{\m}{{\bf m}}

\newcommand{\res}{\upharpoonright}
\newcommand{\Hp}{{\cal H}_{\rm photon}}
\newcommand{\Hgf}{{\cal H}_{\rm gauge field}}
\newcommand{\Uc}{U_{\chi}}
\newcommand{\Uuc}{U^{\chi}}
\newcommand{\Ul}{U_{\lambda}}
\newcommand{\bou}{{\bf u}}
\newcommand{\boe}{{\bf e}}
\newcommand{\Us}{U_{\sigma}}
\newcommand{\Uul}{U^{\lambda}}
\newcommand{\Hc}{{\cal H}_{\chi}}
\newcommand{\Hs}{{\cal H}_{\sigma}}
\newcommand{\Huc}{{\cal H}^{\chi}}
\newcommand{\Hl}{{\cal H}_{\lambda}}

\newcommand{\Hul}{{\cal H}^{\lambda}}
\newcommand{\Lt}{L_{\til{p}}}
\newcommand{\rt}{{\Bbb R}^3}
\newcommand{\rf}{{\Bbb R}^4}
\newcommand{\cf}{{\Bbb C}^4}
\newcommand{\am}{A_{\mu}}
\newcommand{\an}{A_{\nu}}

\newcommand{\bA}{{\bf A}}
\newcommand{\pa}{\partial}
\newcommand{\la}{\langle}
\newcommand{\ra}{\rangle}
\newcommand{\g}{{\bf g}}
\newcommand{\rep}{representation}
\newcommand{\rab}{\rangle_{\frak B}}
\newcommand{\MW}{Marsden-Weinstein}
\newcommand{\Uph}{U_{\rm photon}}
\newcommand{\bol}{{\bf l}}
\newcommand{\bJ}{{\bf J}}
\newcommand{\bK}{{\bf K}}

\newcommand{\Ugf}{U_{\rm gauge field}}
\newcommand{\lt}{L^2({\Bbb R}^3,d'p)\otimes {\Bbb C}^4}
\renewcommand{\pt}{\tilde{p}}
\newcommand{\W}{{\cal W}}
\renewcommand{\O}{{\cal O}}
\begin{document}
\setlength{\baselineskip}{1.5\baselineskip}
\thispagestyle{empty}
\title{Massless Particles, Electromagnetism, \\ and Rieffel Induction}
\author{ N.P.~Landsman\thanks{Alexander von Humboldt Fellow and
S.E.R.C. Advanced Research Fellow; permanent address: Department of Applied
Mathematics and Theoretical Physics, University of Cambridge, Silver Street,
Cambridge CB3 9EW, United Kingdom}\mbox{ } and U.A. Wiedemann\thanks{Ribbands
Scholar, Wolfson
College, Cambridge; supported by an EC Research Training Fellowship under
contract no.\ ERBCHBICT920018; Address from
1-10-1994: Institut f.\ Theoretische Physik, Universit\"{a}t Regensburg,
D-93040  Regensburg, F.R. Germany}\\
\mbox{}\hfill \\ II. Institut f\"{u}r Theoretische Physik, Universit\"{a}t
Hamburg\\ Luruper Chaussee
149, 22761 Hamburg, F.R.\ Germany }  \maketitle \begin{abstract} The connection
between space-time
covariant \rep s (obtained by inducing from the Lorentz group) and irreducible
unitary \rep s (induced
from Wigner's little group) of the Poincar\'{e} group  is re-examined in the
massless case. In the
situation relevant to physics, it is found that these are related by
Marsden-Weinstein reduction with
respect to a gauge group. An analogous phenomenon is observed for   classical
massless relativistic
particles.  This symplectic reduction procedure can be (`second') quantized
using a generalization of
the Rieffel induction technique in operator algebra theory, which
  is carried through in detail for electromagnetism.

Starting from the so-called Fermi \rep\ of
the field algebra generated by the free abelian gauge field, we construct a new
(`rigged')
sesquilinear form on the \rep\ space, which is positive semi-definite, and
given in terms of  a
Gaussian weak distribution (promeasure) on the gauge group (taken to be a
Hilbert Lie group).
This eventually constructs   the algebra of observables of quantum
electromagnetism
(directly in its  vacuum \rep)
as a \rep\ of the so-called algebra of weak observables induced by the trivial
\rep\ of the gauge
group.
 \end{abstract}
 \newpage
\section{Introduction}
This paper is mainly  concerned with the theory of the free electromagnetic
field. Our reason for studying this system
is that it provides the simplest physically relevant model on which to test
certain new ideas to
handle field theories with constraints.
These ideas equally well apply to interacting gauge theories,
and to some extent even to general relativity, so we hope that our formalism
will turn out to be useful in
the quantization theory of those theories, too.

The essential attribute of a gauge theory is that its equations of motion are
simultaneously under-
and overdetermined: the time-evolution of certain components of the gauge field
 is not specified at
all, whereas there are constraints on the Cauchy data. For example, in
classical electromagnetism
(CEM) the Maxwell equations $\square \am -\partial_{\mu} \partial^{\nu}\an=0$
split up (on the choice
of a Cauchy surface $x^0=t={\rm const.}$) into the evolution equations $\square
 \bA^T=0$ and the
constraint $\dot{A}^L-A_0=0$ (Gauss' law); the evolution of
$A_0$ and $ A^L$ is left undetermined (here $\am=(A_0,\bA^T +\nabla A^L )$,
with
$A^L= \Delta^{-1}\nabla\cdot \bA$). It is a remarkable (yet little-known)
feature
of classical gauge theories that the double procedure of imposing the
constraints and factoring out
the remaining undetermined (and unphysical) fields is equivalent to a so-called
Marsden-Weinstein
reduction of the phase space of the unconstrained theory with respect to the
gauge
group \ci{GIMM,BFS}, obviating the need to explicitly perform the above split.
Our  main purpose
 is to exploit this  feature in setting up  a corresponding quantum theory.

We briefly recall this reduction procedure \ci{MW,Mey,AM,Mar74}. Let a Lie
group $G$ act smoothly on a
symplectic manifold $S$; $G$ and $S$ may be infinite-dimensional.
It is required that the action is strongly Hamiltonian. This means firstly that
the
symplectic form $\om$ is invariant under the group action, and secondly that
for each $X\in\g$ (where
$\g$ is the Lie algebra of $G$) one can find a function $J_X$ satisfying
$i_{\til{X}}\om=dJ_X$ (where
$\til{X}$ is the vector field on $S$ defined by the infinitesimal action of
$X$, that is,
$(\til{X}f)(s)=d/dt f(\exp(tX)s)_{|t=0}$), and $\{J_X,J_Y\}=J_{[X,Y]}$, where
$\{\; ,\;\}$ is the
Poisson bracket on $\cin(S)$ derived from the symplectic form, and $[\; , \; ]$
is the Lie bracket on
$\g$. In that case, one can define a so-called moment map $J:S\raw \g^*$ (where
$\g^*$ is the
topological dual of $\g$) by $\la J(s),X\ra =J_X(s)$, which intertwines the
given $G$-action on $S$
and the co-adjoint action on $\g^*$. Then the Marsden-Weinstein reduced space
$S^0$ is defined as
$J^{-1}(0)/G$, which is a symplectic manifold provided  certain technical
conditions are satisfied.

In order to have an optimal setup for quantizing the theory, as well as to
exploit the connection
between electromagnetism and the representation theory of the Poincar\'{e}
group $P$, we depart from
the approach in \ci{GIMM,BFS} in our choice of the classical phase space $S$.
Namely, we would like
the quantum theory to be meaningful in the context of algebraic quantum field
theory \ci{Haa,Hor},
and for this the time-evolution of all fields should be specified, and an
action of $P$ be defined on
them. This necessitates a partial gauge fixing, which we impose by stipulating
that the unphysical
fields $A_0$ and $\bA^L$ satisfy the same equation of motion as $\bA^T$, so
that $\square \am =0$.
The Gauss' law constraint may then be rewritten as $\pa^{\mu} \am =0$.
This procedure is not quite the same as imposing the
Lorentz gauge, for we treat the constraint $\pa^{\mu} \am =0$ as Gauss' law
rather than as a gauge
condition; it is not identically satisfied by the field $\am$.
It may equivalently be arrived at by manipulating the Lagrangian, cf.\ \ci[p.\
143]{Sun}).

We then take $S$ to be the Hilbert space of those real weak solutions $\am$ of
the wave equation $\square \am=0$ whose
Cauchy data lie in $L^2$ in a suitable sense (cf.\ subsection 2.3). This space
is equipped with a symplectic form defined by
\be
\om(B,C)= -\int
d^3x \, B_{\nu}(x) \stackrel{\leftrightarrow}{\partial_{0}}C^{\nu}(x),
 \ll{space}
\ee
For the (residual) gauge group $G$ we choose the Hilbert Lie group of scalar
solutions $\lm$
(modulo constants) of the wave equation whose (exterior) derivative $d\lm$ lies
in $S$. This group
acts on $S$ in the usual way by gauge transformations, i.e., $\lm\in G$ maps
$\am\in S$ to
$\am+\pa_{\mu} \lm$. We will verify that all conditions for Marsden-Weinstein
reduction are
satisfied, so that the reduced    space $S^0$ as defined above indeed coincides
with the physical
phase space of electromagnetism. The main point here is that the constraint
space $J^{-1}(0)\subset
S$   coincides with the space of fields satisfying
 $\pa^{\mu} \am =0$.

{}From the point of view of the Poincar\'{e} group, what happens here is the
following.
Though interpreted above as classical phase spaces of a field theory, $S$ and
$S^0$ are
simultaneously Hilbert spaces of quantum states, which may be construed as the
respective
quantizations of certain one-particle phase spaces (the one corresponding to
$S^0$ being a
co-adjoint orbit of $P$).
The natural action of $P$ on $S$ is  non-unitary, reducible, and
indecomposable if one regards $S$ as a Hilbert space.
But if one looks at $S$ as a symplectic space, this action is symplectic, and
intertwines with the action
of the gauge group in such a way that it has a well-defined quotient action on
$S^0$. This reduced
action is exactly the  irreducible unitary \rep\ defined by massless particles
of helicity $\pm 1$.
It is clear from the indecomposability of $S$ under $P$ that direct integral
decompositions could
not have achieved this reduction.

We now turn to  the (`second') quantization of $S$ (from an algebraic point of
view, the object that is quantized is actually the Poisson algebra
$\cin(S)$). Our main problem is  finding a quantum analogue of the
Marsden-Weinstein
reduction procedure.
The most convenient setting for doing this is provided by algebraic quantum
field
theory  \ci{Haa,Hor}, which has already been extensively used in the study of
quantum electromagnetism
(QEM) \ci{CGH1,CGH2,GH1,GH2,GHJMP,NT}. As explained in the Introduction, this
theory can only be used
if the gauge is partially fixed. Hence we follow \ci{CGH1} in taking the field
algebra $\F$ of QEM
to be the CCR algebra over the symplectic space $S$ (cf.\ \ci{BR2}).

The elimination of the unphysical degrees of freedom in $\F$  was accomplished
in \ci{GH1,GH2}
at a purely algebraic level using the so-called $T$-procedure developed there.
This technique is
intended to provide a rigorous version of the Dirac method of dealing with
quantum constrained
systems \ci{Dir,Sun}. In this paper we propose to handle quantum
electromagnetism using a completely
different method, which has specifically been designed to quantize systems
whose classical constraint
structure is given by \MW\ reduction. This method is based on the observation
\ci{NPL} that a
generalization of \MW\ reduction, in which the strongly Hamiltonian group
action on $S$ is replaced
by a homomorphism of an arbitrary Poisson algebra into $\cin(S)$
\ci{MiW,Xu,NPL}, can be quantized by
the operator-algebraic technique of Rieffel induction \ci{Rie,FD}.

 The aim of this induction technique is to find a
\rep\ $\pi^0$ of a (pre-) $C^*$-algebra $\A$ on a Hilbert space $\H^0$, given a
\rep\ $\pi_0$ of some
other (pre-) $C^*$-algebra $\B$ on a Hilbert space $\H_0$. To accomplish this,
the original
formulation started from a left $\A$- and right $\B$-module $L$.
That is, $L$ is merely a linear space, and a homomorphism of $\A$ into the
algebra ${\cal L}(L)$ of all linear maps on $L$ is given, as well as an
anti-homomorphism of $\B$ into ${\cal L}(L)$. It is not required that the
actions of $\A$ and $\B$ commute (although they do so in our applications).
The key ingredient is then
  a so-called rigging map $\la
\cdot,\cdot \ra_{\B}:L\times L\raw \B$. The latter is a sesquilinear form
(conventionally
assumed linear in the second entry) taking values in $\B$, with the additional
properties holding for
all $\Psi, \Phi\in L$: $$\la\Psi,\Phi\ra_{\B}^*=\la\Phi,\Psi\ra_{\B}; \:\:\:\:
\la \Psi,\Phi
B\ra_{\B}=\la\Psi,\Phi\ra_{\B}B$$ for all $B\in\B$, and
 $$ \la A\Psi,\Phi\rab=\la\Psi,A^*\Phi\rab$$ for all $A\in\A$.
Positivity, in the sense that  $\la\Psi,\Psi\rab\geq 0$ for all $\Psi\in L$,
was required in \ci{Rie},
but dropped in \ci{FD} in favour of the weaker property
$\pi_0(\la\Psi,\Psi\rab)\geq 0$, which
suffices for the induction procedure, for it secures that $\H^0$ is a Hilbert
space.
Moreover, the bound $$\pi_0(\la A\Psi,A\Psi\rab) \leq\,  \n A \n^2
\pi_0(\la\Psi,\Psi\rab)$$ is required
to hold for all $A\in\A$ and $\Psi\in L$; it guarantees that $\pi^0(\A)$ is a
(pre) $C^*$-algebra.
The induced space $\H^0$ is then constructed by first forming
 the algebraic tensor product $L\ot\H_0$, and
 endowing it with a bilinear form $(\cdot ,\cdot)_0$, defined by
 $$
(\Psi\ot v,\Phi\ot w)_0=(\pi_0(\la \Phi,\Psi\rab)v,w),$$
where $(\cdot ,\cdot)$ is the inner product in $\H_0$ (taken linear in the
first entry, unlike
the rigging map). This form is positive semi-definite. Secondly, one builds the
quotient of
$L\ot\H_0$ by its subspace of vectors with vanishing $(\cdot ,\cdot)_0$ norm,
and completes the quotient (equipped with the form inherited from
 $(\cdot ,\cdot)_0$) into a Hilbert space $\H^0$.
Denoting the image of an elementary vector $\Psi\ot v\in L\ot\H_0$ in  $\H^0$
under the projection map
onto the quotient by $\Psi\til{\ot} v$, the representation $\pi^0(\A)$ is then
defined on the
subspace of $\H^0$ of finite linear combinations of such images by $
\pi^0(A)  (\Psi \til{\ot})  v = (A\Psi) \til{\ot}  v  $. By the above bound,
this can be extended
to both $\H^0$ and (if $\A$ is not complete) to the completion of $\A$ by
continuity.

It was shown in \ci{NPL} that the special case of quantum \MW\ reduction with
respect to a locally
compact unimodular group $G$, acting continuously and properly on $S$, is
covered as follows. We
assume that the Poisson algebra $\cin(S)$ has been quantized into a
$C^*$-algebra $\F$, which is
faithfully represented  on a Hilbert space $\H$. This space should additionally
carry a continuous
unitary \rep\ $U$ of $G$ (which plays the r\^{o}le of the quantization of the
$G$-action on $S$). Then
$\ovl{\B}=C^*(G)$ acts from the right (that is, in an anti-\rep) on $\H$ by
$\pi^-(f)=\int_G dx\,
f(x)U(x^{-1})$, where $dx$ is the Haar measure on $G$ (this expression is
defined on $C_c(G)$ and
extended to $C^*(G)$ by continuity).  We then try to identify a dense subspace
$L\subset \H$, such
that the integral $\int_G dx\, (U(x)\Psi,\Phi)$ is finite for all $\Psi,\Phi\in
L$ (for $G$ compact
one may take $L=\H$; in general, many choices of $L$ may exist, or none at
all). If so, the rigging
map defined by $\la\Psi,\Phi\rab:x\raw (U(x)\Phi,\Psi)$   takes values in the
pre-$C^*$-algebra
$C(G)\cap L^1(G)$ (whose completion is $C^*(G)$).
We now define
  the so-called  weak algebra of observables
$\F^G$, which is  the subalgebra of operators in $\F$ which commute with all
$U(x)$.
If we take for $\A$ a suitable dense subalgebra
of $\F^G$, then all
properties required of the rigging map are satisfied.

 The fact that our \MW\
reduction is from the value $0\in \g^*$ is then reflected by our taking the
trivial \rep\ of $C^*(G)$
(or $G$) on $\H_0={\Bbb C}$ to induce from; hence the rigged inner product is
defined on $L$ itself,
and is given by
\be
 (\Psi,\Phi)_0=\int_G dx\, (U(x)\Psi,\Phi). \ll{1.2}
\ee
(Thus for $G$ compact, a case which even the naive Dirac formalism \ci{Dir,Sun}
can handle, one simply has
$(\Psi,\Phi)_0=(P_0\Psi,P_0\Phi)$, where $P_0$ projects on the subspace of $\H$
carrying the trivial
\rep\ of $G$).

It should now be clear which problems we face if we wish to apply this scenario
to QEM. The gauge
group $G$ is not locally compact (unless we equip it with the discrete
topology, which would be
disastrous both for \MW\ reduction,  as this procedure is based on the use of
Lie groups, as well as
for Rieffel induction, for more subtle reasons to become clear in section 4),
so it has no Haar
measure. This  means that the group algebra $C^*(G)$ is not defined, and also
 that
(\ref{1.2}) above makes no sense. (See \ci{Gru} for a definition of a group
algebra of groups which
are a topological inductive limit of locally compact subgroups. It remains to
be seen whether this is
of any help for our problem.) Hence we have to rethink the Rieffel induction
procedure, and in doing
so it becomes clear that one neither needs the algebra  $\B$ nor the rigging
map: the essential point
is the rigged inner product $(\cdot,\cdot)_0$ on $L\ot \H_0$ (which coincides
with $L$ in our case,
where $\H_0=\Bbb C$; in what follows we specialize to this case). It suffices
to find a sesquilinear
form on $L$ with the following properties to hold for all $\Psi,\Phi\in L$ and
all $A\in \A$:
\be
1.\:\:\: (\Psi,\Psi)_0\geq 0; \ll{P1}
\ee
\be
2. \:\:\: (A\Psi,\Phi)_0=(\Psi,A^*\Phi)_0; \ll{P2}
\ee
 \be 3.\:\:\:
(A\Psi,A\Psi)_0 \leq \, \n A\n ^2(\Psi,\Psi)_0. \ll{P3}
\ee
The second property implies  that  $\A$ maps the null
space of $\rip$ into itself.
The induced space $\H^0$ and the induced \rep\ $\pi^0(\A)$  are then defined as
before,
 viz.\  $\pi^0(A)\til{\Psi}=\widetilde{A\Psi}$, where
$\til{\Psi}$ is the image of $\Psi$ in the quotient of $L$ by the null space.
 Now (\ref{P1}) guarantees that $\H^0$ is a Hilbert space, and
(\ref{P2}) implies that $\pi^0$ is a $\mbox{}^*$-\rep\ of $\A$, which by
(\ref{P3}) is continuous,
and
 extendable to the completion
$\ovl{\A}$.

Clearly,  (\ref{P1})-(\ref{P3}) are merely  the conditions which
a positive semi-definite sesquilinear form on a left $\A$-module has to
satisfy in order to produce a \rep\ of (the completion of) $\A$. As such, these
conditions  have nothing to do with Rieffel induction. The point is that
Rieffel induction, or our
slight bending of it, provides a systematic mechanism leading to such a form.
Indeed, on the basis of the connection between symplectic reduction and
Rieffel induction \ci{NPL} it may be said that \MW\ reduction itself
provides this mechanism.

We will take $\H=\exp(S)$, the bosonic Fock space \ci{BR2}
 (alternatively known as symmetric Hilbert space \ci{Gui}) over $S$, which
carries a natural \rep\ of $\F$ \ci{CGH1}, and in addition a unitary \rep\ $U$
of the gauge group
$G$.  Inspired by (\ref{1.2}), we attempt to construct the rigged inner product
by
 \be
(\Psi,\Phi)_0=
\lim_n \int_{\H_n} \frac{d^nx}{\pi^{n/2}}\,(U(x) \Psi,\Phi). \ll{1.3}
\ee
 Here $\{\H_n\subset G\}_n$ is an inductive  family of Hilbert subspaces of
$G$, which eventually
exhaust it, and   $d^nx$ is the Lebesgue measure on each  $\H_n\simeq {\Bbb
R}^n$.
Remarkably, this limit indeed exists for a suitable choice of $L$, and can be
written as a
functional integral over $G$ with respect to a Gaussian promeasure (cf.\
\ci{AMP}; this is also
called a weak distribution \ci{Sok}). Moreover, the expression (\ref{1.3})
allows us to prove
properties 1-3 above quite easily (based on the corresponding proofs for the
locally compact case
\ci{NPL}). The induced   \rep\ $\pi^0$ on $\H^0$ can be identified explicitly.
For
example, the Poincar\'{e} automorphisms on $\A$ are implemented in $\pi^0$.

The resulting structure may be compared with the Gupta-Bleuler (or BRST)
formalism of QEM
\ci{SW,GHJMP}.  There one has a \rep\ of the field algebra $\F$ on a space with
indefinite metric,
which contains a non-dense subspace on which the metric is positive
semi-definite. Our form
$(\cdot,\cdot)_0$, on the other hand, enjoys the latter property  on  a dense
subspace $L$. The need
to subsequently quotient out the null space  arises in both formalisms.  Hence
in our formalism there
is no need to first identify a `physical' subspace of $\H$, and we have no
negative norm states. The
price to be paid for this is that  $\H$ does not carry a unitary \rep\ of the
Poincar\'{e} group
\ci{CGH1}, and that the rigged inner product is only densely defined (in fact,
as a quadratic form on
$\H$ it is not even closable). One still has the Hilbert inner product
$(\cdot,\cdot)$ on $\H$, but
it is only used to construct the rigged inner product $(\cdot,\cdot)_0$, and
plays no independent
r\^{o}le whatsoever; the physically  relevant Hilbert inner
product is the one on the induced space $\H^0$, which comes from
$(\cdot,\cdot)_0$ rather than
$(\cdot,\cdot)$.  The definition of the rigged inner product
 via an auxiliary Euclidean structure is responsible for  the fact
that the first step of our procedure is not Lorentz invariant; full
Poincar\'{e} invariance
is restored only on the induced space $\H^0$. It may be possible to
construct $L$ and $\rip$ in a different way, in which
 Poincar\'{e} invariance is manifest at all stages.

 More generally, the usual first step in the quantization of constrained
systems of
first finding a Hilbert space \rep\ of the unconstrained systems should be
replaced by the problem
of finding a \rep\ on a space $L$ carrying a rigged inner product with the
properties 1-3 listed
above, where the weak algebra of observables
 $\ovl{\A}$ is the subalgebra of operators of the
unconstrained system which commute with the constraints, or with the action of
the gauge group.
The actual algebra of observables of the physical system in question is then
$\pi^0(\ovl{\A})$,
which is isomorphic to the quotient of $\ovl{\A}$ by the kernel of $\pi^0$
(note that $\ovl{\A}$, unlike $\F$,
is not simple in general).

To close this Introduction, we briefly summarize the contents and the logic of
this paper.
 As a preliminary, subsection 2.1 contains a brief review of the (non-unitary)
`covariant' (i.e., vector, tensor, etc.) \rep s of the Poincar\'{e} group $P$,
and their connection
with the irreducible unitary (`canonical') \rep s. This leads into
  subsection 2.3, which contains a
  theorem stating that (unitary) massless helicity 1 or 2 \rep s are obtained
as \MW\ reductions of
canonical vector or
 tensor \rep s. An important ingredient of the proof is isolated in a
finite-dimensional model in
subsection 2.2.
This result is central to the paper, for it provides the specific way of
writing
electromagnetism as a constrained system, that we are going to quantize
systematically with Rieffel
induction. In particular, the precise r\^{o}le of gauge transformations and
gauge invariance in the
classical theory is now formulated in such a way as to admit a quantization
from   first
principles.
 As an aside,
we show in subsection 2.4 how a similar theory of covariant versus canonical
realizations of the
Poisson algebra defined by $P$ can be formulated at the classical one-particle
level; here the
r\^{o}le played by massless particles is seen to be very similar to the Hilbert
space case.

As a preparation for our field-theoretic calculations, section 3  is devoted to
a model with four
degrees of freedom (mimicking the components of $\am$ evaluated at a fixed
point in momentum space).
Performing Rieffel induction on this model already exhibits most of the
combinatorial features of the
full theory, while remaining straightforward analytically. This model has the
special feature that, on
account of the Stone-von Neumann theorem, the unconstrained system admits a
unique quantization
(in the sense of an irreducible \rep\ of the algebra generated by its degrees
of freedom).
In that case our construction becomes a strict algorithm - the only freedom
left is that of choosing
between various unitarily equivalent realizations. The key brickwork is the
rigged inner product in
subsection 3.2, which may be thought of as providing a suitable `dual' harmonic
decomposition of the
delta-function on the dual of the gauge group as an integral over the gauge
group. (Recall that in
ordinary harmonic analysis one expresses the delta-function on the group,
rather than its dual, as a
Plancherel integral over the unitary dual of the group.) From this, the induced
\rep\ of the algebra
of observables is derived in a straightforward fashion in subsection 3.3.

 In section 4 we turn to full electromagnetism. Being infinite-dimensional,
   the unconstrained system now no longer has a unique quantization (that is,
an irreducible \rep\ of
the field algebra), and in this sense our construction hinges on making a
certain choice.
 We perform Rieffel
induction on the Fermi \rep\ of the field algebra of quantum electromagnetism,
for this is the
\rep\ that most closely corresponds to the (unique) one used in the
finite-dimensional model in
section 3. Once this choice has been made, the rigged inner product (in
subsection 4.2) is
`canonical'; it has the same interpretation as the one in the preceding
section, with the new feature
that our `dual' harmonic analysis now involves a (rigorous) functional
integral.  We construct the
corresponding algebra of observables in its vacuum \rep\ in subsection 4.3.

 We close with some loose remarks in section 5. The main part is subsection
5.3, where we comment on
the subtle differences and analogies between our method and the $T$-procedure
of Grundling and Hurst.

We wish to acknowledge our debt to the paper \ci{CGH1}, which introduced the
models we discuss in
sections 3 and 4, and, using different techniques, analyzed many of their
quantum-mechanical  features.
Helpful discussions with Hendrik Grundling took place at an early stage of this
work, as well
as after completion of the first draft.  Rainer Verch provided us with
constructive criticism of the
manuscript. Also, thanks to Bernard Kay for discussions on eq.\
(\ref{arakieq}), and pointing out
ref.\ \ci{Ara}.

 Our metric is $g={\rm diag}\, (1,-1,-1,-1)$, Greek indices run from 0 to 3,
and Latin ones from 1
to 3. If no confusion can arise, we omit the tilde or hat on
Fourier-transformed functions.
The Lie algebra of a Lie group $G$ is denoted by $\g$.

\section{Classical gauge theories and the representation theory of the
Poincar\'{e} group}\eo
The aim of this section is to show that the passage from the \rep\ of the
Poincar\'{e} group $P$
carried by the massless vector field $\am$ to the unitary irreducible \rep\
defined by massless
particles of helicity $\pm 1$ can be accomplished by \MW\ reduction.
\subsection{Review of covariant vs.\ canonical \rep s}
 References for this subsection are  \ci{Wei,Car,BR}.
 The  Poincar\'{e} group is the regular semidirect product $P=L \ltimes N$,
where $L$ is Lorentz
group and  $N=\R^4$ (we are confident that the reader will not confuse this
notation with the linear
space $L$ in Rieffel induction). The Mackey theory of induced \rep s of regular
semidirect products
applies to a general abelian factor $N$ and locally compact group $L$,  but
since time inversion is
represented by an anti-linear operator in physics, we can only use this theory
for the proper
orthochronous subgroup $P_+^{\uparrow}$ of $P$. Hence, in what follows   $L$
stands for
$L_+^{\uparrow}$ (i.e., $\det \Lm =1$ and $\Lm^0_{\mbox{ }0}\geq 0 \:\forall
\Lm\in L$).

Canonical and covariant \rep s are both examples of induced \rep s. That is,
take a closed subgroup
$H\subset P$ and a \rep\ (not necessarily unitary) $\Uc$ of $H$ on a Hilbert
space $\Hc$. Then choose
a measurable
section $s:P/H\raw P$ of the canonical projection $pr_P:P\raw P/H$
(i.e., $pr_P\circ s=id$). Assume that $P/H$ has a
$P$-invariant measure, w.r.t.\ which we define $L^2(P/H)$. The induced \rep\
$\Uuc$ of $P$ on
the Hilbert space $\Huc=L^2(P/H)\ot \Hc$ is then given by
\be
(\Uuc(x)\ps)(q)=\Uc(s(q)^{-1}\, x\, s(x^{-1}q))\ps(x^{-1}q). \ll{2.1}
\ee
This is unitary iff $\Uc$ is unitary. The two relevant specializations of this
scheme are:

{\em 1. Canonical \rep s.} $L$ acts on $\hat{N}$ (the dual group of $N$; for
$N=\R^4$, $\hat{N}=\R^4$
and we take the Minkowski pairing between $N$ and $\hat{N}$, i.e., $\langle
p,a\rangle=
\exp(ip_{\mu}a^{\mu})$) by $\langle \Lm p,a\rangle =\langle
p,\Lm^{-1}a\rangle$.
Take $\til{p}\in \hat{N}$ fixed and let $L_{\til{p}}$ be the stability group of
$\til{p}$.
Then in the above scheme $H=\Lt\ltimes N$, and $P/H\simeq L/ \Lt$.
If we choose a section $\et: L/\Lt\raw L$ of
the canonical projection $pr_L\equiv pr:L\raw L/\Lt$, we have also
obtained a section $s: L/ \Lt\raw P$, given  by
$s(q)=(\et(q),0)$. A \rep\ $U_{\sg}$ of $\Lt$ on a Hilbert space $\H_{\sg}$
defines a
\rep\ $U_{\til{p},\sg}$ of $H$
on $\H_{\til{p},\sg}=\H_{\sg}$ by $ U_{\til{p},\sg} (\Lm,a)=\exp(i\til{p}a)
U_{\sg}(\Lm)$, with $\til{p}a\equiv
 \til{p}_{\mu}a^{\mu}$.
Thus $\H^{\til{p},\sg}=L^2(L/\Lt)\ot \H_{\sg}$. Up to unitary equivalence, the
induced \rep\ defined
on this space only depends on the orbit of $\til{p}$, which for $\til{p}^2\geq
0$ is of the form
$O_{m}^{\pm}=\{p\in\rf|p^2= m^2, \pm p^0>0 \}$. Hence we relabel
$\H^{\til{p},\sg}$ as
$\H^{m,\pm,\sg}$;
 the
induced \rep\ follows from (\ref{2.1}) as
 \be
 (U^{m,\pm,\sg}((\Lm,a))\ps)(p)=e^{ipa} U_{\sg}(\et(p)^{-1}\,
\Lm\, \et(\Lm^{-1}p))\ps(\Lm^{-1}p). \ll{2.2}
 \ee
To emphasize that $L/\Lt$ is identified with the $L-$orbit $O_{\til{p}}$ of
$\til{p}$ in
$\R^4$, we have denoted its points by $p$ rather than $q$. It is well known
that the
unitarity and irreducibility of $U^{m,\pm,\sg}$ is implied by the corresponding
properties of
$U_{\sg}$.

{\em 2. Covariant \rep s.}
Here $H=L$, so that $P/H\simeq \R^4$. Now we induce from a finite-dimensional
\rep\ $\Ul$ of
$L$ on $\Hl$ (which is never unitary unless it is the trivial \rep), and
conveniently choose
$s(x)=({\rm id},x)$. Hence on $\Hul=L^2(\R^4)\ot \Hl$ the induced \rep\ given
by (\ref{2.1}) becomes
\be
(\Uul(\Lm,a)\ps)(x)=\Ul(\Lm)\ps(\Lm^{-1}(x-a)), \ll{2.3}
 \ee
or, after a Fourier transform $\hat{\ps}(p)=(2\pi)^{-1}\int d^4x\,
\ps(x)\exp(ipx)$,
\be
(\Uul(\Lm,a)\hat{\ps} )(p)=e^{ipa}\Ul(\Lm)\hat{\ps}(\Lm^{-1}p). \ll{2.4}
\ee
The first step in the reduction of this highly reducible \rep\ is to decompose
it as a direct
integral over the orbits $O_{\pm m}=\{p\in\rf|p^2=\pm m^2\}$ in $\hat{N}$. We
take the + sign only
in what follows, and must then further decompose over the two orbits
$O_{m}^{\pm}$. This leads to the
Hilbert spaces $\hat{\H}^{m,\pm,\lm} =L^2(\R^3,d'p) \ot \Hl$   with the
invariant measure
$d'p=d^3p/(2\pi)^32 \sqrt{\bop^2+m^2}$.
  This space carries a \rep\
$\hat{U}^{m,\pm,\lm} $, which is the (improper) restriction of $\Uul$, and is
therefore simply given
by putting $p^0=\pm \sqrt{\bop^2+m^2}$ in
(\ref{2.4}).

The connection with the canonical \rep s discussed before follows by
 picking a point $\til{p}\in O^{\pm}_{m}\simeq L/\Lt$, and a section $\et$,
as before,  and defining the transformation
$V$ on $\hat{\H}^{m,\pm, \lm}$ by
\be
\ps(p)\equiv (V\hat{\ps})(p)=\Ul(\et(p)^{-1})\hat{\ps}(p). \ll{2.5}
\ee
If  we define $\H^{m,\pm, \lm}_{\et}$ to coincide with $\hat{\H}^{m,\pm, \lm}$
as a vector space, but
equipped with the modified inner product
\be
(\ps,\phv)_{\et}=\int d'p
(\Ul(\et(p)^{-1})\ps(p),\Ul(\et(p)^{-1})\phv(p))_{\H_{\lm}}, \ll{ip}
\ee
and refer to   $\hat{\H}^{m,\pm, \lm}$ with its usual inner product as
$\H^{m,\pm, \lm}$,
then $V:\H^{m,\pm, \lm}_{\et}\raw \H^{m,\pm, \lm}$ is evidently unitary.
The \rep\ $\hat{U}^{m,\pm,\lm} $, but now defined on $\H^{m,\pm, \lm}_{\et}$
rather than
$\hat{\H}^{m,\pm, \lm}$, is relabeled as $ U_{\et}^{m,\pm ,\lm} $.
 The representation $ U^{m,\pm ,\lm} =V  U_{\et}^{m,\pm ,\lm} V^{-1}$ is
then given by
\be
( U^{m,\pm ,\lm}  ((\Lm,a))\ps)(p)=e^{ipa} U_{\lm}(\et(p)^{-1}\, \Lm\,
\et(\Lm^{-1}p))\ps(\Lm^{-1}p). \ll{2.6}
\ee
That is, it coincides with (\ref{2.2}) up to the fact that (\ref{2.6}) contains
$\Ul(\Lt)$ on $\Hl$
(i.e., the restriction of $\Ul$ from $L$ to its subgroup $\Lt$), whereas
(\ref{2.2}) has the \rep\
$\Us(\Lt)$ on $\Hs$ (which we assume to be unitary and irreducible). Since
$\Ul(\Lt)$ is generically
reducible, so is $ U^{m,\pm, \lm} $. To recap: the pair $(\hat{U}^{m,\pm
,\lm},\hat{\H}^{m,\pm ,\lm})
=(\hat{U}^{m,\pm ,\lm},\H^{m,\pm ,\lm})$ is
independent of $\et$, but irrelevant. In $(U_{\et}^{m,\pm ,\lm},\H_{\et}^{m,\pm
,\lm})$ only
$\H_{\et}^{m,\pm ,\lm}$ explicitly depends on
$\et$ through its inner product. This pair is unitarily equivalent to
$(U^{m,\pm ,\lm},\H^{m,\pm ,\lm}
)$, in which only $U^{m,\pm ,\lm}$
depends on $\et$.

For $m>0$ the reduction of  $\Ul(\Lt)$ is straightforward, because $\Lt=SO(3)$
is
compact, so that  $\Ul(\Lt)$ is completely reducible, and the desired
irreducible component can be
projected out by further covariant subsidiary conditions.
 For $m=0$, on the other hand,  the potential problem arises
that $\Lt=E(2)=SO(2)\ltimes \R^2$, which is neither semi-simple nor compact,
has indecomposable \rep
s. However,   the representation $D_{(j_1,j_2)}$ of $L$, restricted to $E(2)$,
contains a
subrepresentation of helicity $h$ iff $h=j_2-j_1$. Hence the tensor
$\til{F}_{\mu\nu}$, which
carries the \rep\ $D_{(1,0)}\oplus D_{(0,1)}$, is suitable to describe the
helicity $\pm 1$
representations of the physical photon; the free Maxwell equations merely
project these \rep s out of
the reducible $F_{\mu\nu}$ (note that the tilde on $F_{\mu\nu}$ is written to
indicate that
physically $\til{F}_{\mu\nu}=F_{\mu\nu}(\pt)$, see below).

By the same token, the  vector $\til{A}_{\mu}$ carrying the \rep\
$D_{(\half,\half)}$ of $L$
(which is defined by $D_{(\half,\half)}(\Lm) \til{A}_{\mu}=\Lm_{\mu}^{\mbox{
}\nu}\til{A}_{\nu}$),  does
not contain these sub\rep s.
The reason is well-known, but it is worthwhile to recall it, and reformulate
the ensuing discussion
in the language of symplectic reduction theory.
 \subsection{Marsden-Weinstein reduction for the frozen photon field}
We conventionally choose the fixed
point  $\pt\in O_0^+$ to be $\pt^{\mu}=(1,0,0,1)$, and label
  the complex Fourier coefficients $\am(\pt)$ as $ \til{\am}\in \cf$.
The little group
$\Lt=E(2)$ (consisting of those Lorentz transformations which leave $\pt$
stable) is embedded in $L$
by identifying the abelian generators $\hat{T_1},\, \hat{T_2}$ of the former
with $\hat{K}_1-\hat{J}_2,\,
\hat{K}_2+\hat{J}_1$, respectively, and its generator of $SO(2)$ with
$\hat{J}_3$; here
$\hat{K}_i=\hat{M}_{0i}$ and $\hat{J}_i=\half\ep_{ijk}\hat{M}_{jk}$, where
$\hat{M}_{\mu\nu}=-
\hat{M}_{\nu\mu}$ are the usual generators of the Lie algebra $\bol$ of the
Lorentz group. To describe
how $\H_{(\half,\half)}=\cf$ decomposes under $D_{(\half,\half)} (L\uph E(2))$,
we choose a basis
$\{\bou_1,\bou_2,\bou_+,\bou_-\}$, which is related to the usual basis
$\{\boe_{\mu}\}$ of $\cf$ by
$\bou_1=\boe_1,\, \bou_2=\boe_2,\, \bou_{\pm}=\half(\boe_0\pm\boe_3)$. Since
 $\pt_{\mu}=(1,0,0,-1)$,
 $T\equiv {\Bbb C}\,\bou_-$
is invariant under $E(2)$, and so is $N\equiv {\rm
span}(\bou_1,\bou_2,\bou_-)$, but the latter does
not decompose, since $\hat{T}_i\bou_i=\bou_-$ ($i=1,2$).

The desired \rep\ $U_1\oplus U_{-1}$ of $E(2)$ on $\C^2$, which characterizes
photons, is obtained by
quotienting $N$ by $T$: the natural action of $E(2)$ on the quotient
indeed coincides with $U_1\oplus U_{-1}$ (i.e., $\R^2$ acts trivially and
$SO(2)$ acts in its
(complexified)
defining \rep). Hence one proceeds in two steps: first a constraint
$\pt^{\mu}\til{\am}=0$ is
imposed, and then one factorizes the solution space $N$ of the constraint
by the equivalence relation $\til{\am}\sim \til{\am}+\lm\pt_{\mu}$, $\lm\in\C$.

This procedure may be reformulated as a \MW\ reduction.
 To treat $\cf$ as a classical
phase space $(S,\om)$ we equip it with the symplectic form $\om$ defined by
 \be
\om(\til{B},\til{C})=2\, {\rm Im}\,
(\til{B},\til{C})_M\equiv 2\, {\rm Im}\, \til{B}^{\mu}\ovl{\til{C}}_{\mu}.
\ll{minisym}
\ee
 Then the frozen gauge group
$G=\C$  (regarded as additive group) acts on $S$ in that $\lm\in G$ maps
$\til{A}\in S$ to
 $\til{A}+\lm\pt$.
This action is strongly Hamiltonian, with moment map $J:S\raw \g^*=\C$ given by
$J(\til{A})=i\pt^{\mu}\til{A}_{\mu}$. Hence the constraint set $N$ coincides
with $J^{-1}(0)$, and
the reduced space $\C^2$ constructed above is nothing but the \MW\ quotient
$S^0=J^{-1}(0)/G=N/T$.

We now compute the symplectic form $\om^0$ on $S^0$. Let $\til{A}\in N$, and
$\til{B},\, \til{C}\in
T_{\til{A}}N\simeq N$. We denote the equivalence classes of these vectors in
$N/T$ by
$[\cdot]$. By the theory of \MW\ reduction, $\om^0$ is defined at $[\til{A}]$
by
\be
\om^0([\til{B}],[\til{C}])=\om(\til{B},\til{C}). \ll{om1}
\ee
In general, the right-hand side should be evaluated at any lift $\til{A}$ of
$[\til{A}]$, but in the
present case $\om$ is the same at all points of $S$.
 In any case, by the $G$-invariance of $\om$, the r.h.s.\ is independent of
the chosen representative $\til{\cdot}$   of the class $[\til{\cdot}]$. Hence
if we define an
inner product $(\cdot,\cdot)_{S^0}$ on $S^0$ by
 \be
([\til{B}],[\til{C}])_{S^0}=(\til{B}_T,\til{C}_T)_S, \ll{om2}
\ee
where the inner product on the r.h.s.\ is the Euclidean one in $\cf$, and
$\til{B}_T=(0,\til{B}_1,\til{B}_2,0)$ (guaranteeing independence of the choice
of representatives),
then
\be
\om^0([\til{B}],[\til{C}])  =-2\, {\rm Im}\,
([\til{B}],[\til{C}])_{S^0}. \ll{om3}
\ee
Moreover, the reduced \rep\ of $E(2)$ on $S^0$ is unitary with respect to the
inner product on $S^0$.

Anticipating the quantization of the model in section 3, one may introduce a
classical `field
algebra' ${\cal F}=\cin(\cf)$, whose algebraic structure is given by the
Poisson bracket derived
from $\om$ and by pointwise multiplication. The algebra of weak observables
${\cal A}$ is then
given by ${\cal F}^G$, which stands for the set of those elements $f$ of $\cal
F$ which are
gauge-invariant (that is, satisfying $f(\til{\am}+\lm\pt_{\mu})=f(\til{\am})$
for all $\til{A}\in S$
and $\lm\in G$). The elements of $\cal A$ have a well-defined action on $S^0$,
which may be thought
of as providing a classical induced \rep\ of $\cal A$, cf.\ \ci{NPL}.
The quotient of $\cal A$ by the kernel of this \rep\ is then the algebra of
observables of the model.

  We will now see how this  procedure works for the
entire field.
\subsection{\MW\ reduction for the photon field}
In the
notation we used for canonical \rep s,
 the one-photon Hilbert space is
$ \Hp=\H^{0,+,1}\oplus \H^{0,+,-1}$,
where the labels $\pm 1$ refer to the helicity $\pm 1$ \rep s of $E(2)$. Our
notation implies that
$\Hp$ is to be seen as a $P$-module carrying the unitary \rep\
\be
 \Uph=U^{0,+,1}\oplus U^{0,+,-1} , \ll{uph}
\ee
cf.\ (\ref{2.2}).
  The momentum space gauge field $\am(\bop)$ is taken to transform under the
covariant \rep
\be
 \Ugf= U^{0,+,(\half,\half)}  \ll{ugf}
\ee
 induced by
 $D_{(\half,\half)}$ of $L$, cf.\ (\ref{2.6}) .
Hence
 the gauge field Hilbert space is
$\H_{ \rm gauge field}=   \H^{0,+,(\half,\half)}$, identified with
$\lt$.   The connection between (\ref{ugf}) and (\ref{2.4}) is somewhat subtle.
The space  of real weak solutions
$\am(x)$ of the wave equation $\square \am =0$ with Fourier coefficients (in
the sense that
$\am(x)=\int d'p\, \am(\bop)\exp(-ipx)+c.c.$)
 in
$L^2(\rt,d'p)\ot\cf$ forms a real Hilbert space, and defines a real \rep\ of
$P$ through (\ref{2.3}).
This corresponds to a real sub\rep\ of $U^{0,+,(\half,\half)}\oplus
U^{0,-,(\half,\half)}$, which in
quantum field theory is juggled so as to be replaced by the complex \rep\
(\ref{ugf}). Accordingly, we
have taken the shortcut of starting from  (\ref{ugf}) directly.

Our problem is now to pass from $\Hgf$ to $\Hp$; as we have pointed out, this
is not possible using
a Hilbert space decomposition, but it can be achieved by \MW\ reduction.
For this purpose we equip $\Hgf$ with a symplectic form defined by
\be
\om(B,C)=2\, {\rm Im} \int d'p B_{\mu}(\bop)\ovl{C^{\mu}(\bop)}\equiv 2 {\rm
Im} (B,C)_M, \ll{2.7}
\ee
cf.\ (\ref{minisym}).  In $x$-space,  this form was
already given in (\ref{space}).
We will henceforth refer to  $\Hgf$ with $\om$ as $S$, to stress that it
is now regarded as real  symplectic (Hilbert) manifold.

It is easily shown that $S$ is
strongly symplectic (for one can modify the complex structure of $S$ so that
$\om$ becomes  twice the
imaginary part of the inner product, cf.\ \ci{CGH1}).

We take the gauge group to be the Hilbert space $G=L^2(\rt, d^3p\,
|\bop|/2(2\pi)^3 )$, that is, we
take
  non-constant weak solutions $\lm$ of the wave equation whose
Fourier-transformed exterior
derivative $d\lm$ lies in $S$. The norm is accordingly
\be
\n \lm \n_{\half}=\half(d\lm,d\lm)_S=\left( \int d'p\, |\bop |^2
|\lm(\bop)|^2\right)^{\half}. \ll{Sob}
\ee
The abelian group structure is given by addition, and the gauge group acts on
$S$ by gauge
transformations, i.e., in $x$-space $A\raw A+d\lm$ as usual,
and hence in  momentum space by $\am(\bop)\raw \am(\bop)-ip_{\mu}\lm(\bop)$
(with the
standing notation $p_0=|\bop|$); for brevity we will write the latter action
as $A\raw A+d\lm$ as
well. The topology on $G$ has been chosen precisely so that this action is
smooth.

  This action
preserves the symplectic form
 and is
even strongly Hamiltonian, with moment map $J:S\raw {\rm Lie}(G)^*$  given by
 \be
 J_X(A)\equiv \langle J(A),X\rangle
= \om(dX,A), \ll{mm}
 \ee
  cf.\ (\ref{2.7}).
Thus, if we regard $\g={\rm Lie}(G)\simeq G$ as a real Hilbert space, and
identify $\g^*$ with $\g=G$
under the pairing $\la \th,X\ra=2\, {\rm Im}(X,\th)$, then $J(A)$ is the
element $\hat{A}$ of $G$
whose value at $\bop$ is given by $p^{\mu}\am(\bop)/|\bop|^2$; note that this
function indeed lies in
$L^2(\rt, d^3p\, |\bop|/2(2\pi)^3 )$ if $A$ is in $\lt$. It easily follows that
$J$ is smooth, and
because $J(A+d\lm)(\bop)=J(A)(\bop)-i p^2 \lm(\bop)/|\bop|^2=J(A)(\bop)$ (for
$p^2=0$), it is
clearly equivariant (recall that the co-adjoint action of $G$ on $\g^*$ is
trivial, as $G$ is
abelian). We now show that the \MW\ reduced space of $S$ exists and carries the
desired \rep\ of $P$.
 \begin{theorem}
The \MW\ reduced space $S^0=J^{-1}(0)/G$ is a symplectic Hilbert manifold.
In addition, it is a linear space, which inherits an inner product and a
$P$-action  (see
(\ref{ugf})) from $S$. With respect to these inherited structures, it is
unitarily equivalent to the
Hilbert space $\Hp$ carrying the \rep\ $\Uph$ (cf.\ (\ref{uph})).
\end{theorem}
{\em Proof}.
We first show that $S^0$ is a symplectic manifold. Given that we already know
that $(S, \om)$ is
strongly symplectic, that $J$ is equivariant,  and that $S$ is a Hilbert
manifold (so that, in
particular, its model space is reflexive), this follows from the theory of
(infinite-dimensional)
\MW\ reduction (cf.\ \ci{AM}, and esp.\ \ci[Ch.\ 6]{Mar74}) if: i) 0 is a
regular value of $J$, and
ii) $G$ acts freely and properly on $J^{-1}(0)$.

To show i), we compute the derivative $J_*(A)$ of $J$ at any $A$ to be
$J_*(A):B\in T_AB\raw \hat{B}\in T_{J(A)}\g^*\simeq \g^*\simeq G$ (cf.\ the
preceding par.), which of
course is independent of $A$. Thus an arbitary $\ps\in G$ is the image of
$B_{\mu}(\bop)=\half
p^{\mu} \ps(\bop)$ (note that $p^{\mu}p^{\mu}=2 |\bop|^2$), so that $J_*(A)$ is
surjective at any
$A\in S$, hence certainly at $A\in J^{-1}(0)$.

As to   ii), the freeness of the $G$-action is trivial, whereas the properness
follows from the
equivalent property that $A_n\raw A$ and $\lm_n\cdot A_n\raw B$ in $S$ together
 must imply that
the sequence $\{\lm_n\}$ in $G$  has a convergent subsequence. Since the group
action is given
by $\lm\cdot A=A+d\lm$, this follows immediately from the topology we have put
on $G$
(in which $\lm_n\raw\lm$ in $G$ is the same as $d\lm_n\raw d\lm$ in $S$).

The set $J^{-1}(0)$ consists of those $A$ satisfying $p^{\mu}\am(\bop)=0$ for
almost all $\bop$,
and is closed: if $A_n\in J^{-1}(0)$ and $A_n\raw A$ in $S$, then $A\in
J^{-1}(0)$ because of
(\ref{mm}) and Cauchy-Schwartz.
We now fix a Lorentz frame, and define the projector $P_T$ on $S$ by
\be
 P_T A_0=0;\:\:\: (P_T A_i)(\bop)=A_i(\bop)-p_i p_j A_j(\bop)/|\bop|^2, \ll{pt}
\ee
Thus   $J^{-1}(0)$ has the (non Lorentz-invariant) orthogonal decomposition
$J^{-1}(0)=P_T
S\oplus d G$, where $d:G\raw S$ is given   by
$(d\lm)_{\mu}(\bop)=-ip_{\mu}\lm(\bop)$, cf.\ \ci[p.\
636]{CGH1}. Hence $J^{-1}(0)/G\simeq P_T S$, which shows that $S^0$ is a
Hilbert space.
 (This is {\em not} an isomorphism as carrier spaces of the action of the
Poincar\'{e} group $P$,
which does not leave $P_TS$ stable).

  Furthermore, $P$ acts on $S$ by symplectic transformations, maps $J^{-1}(0)$
into
itself (as the condition defining this space is invariant), and also maps a
vector of the type $d\lm$
into another vector of this type (for $\Ugf d\lm =d \Ugf\lm$), so that it
preserves the
$G$-foliation. The $P$-action on $S$ therefore quotients to an action on $S^0$,
and it follows from
the isomorphism $S^0\simeq P_T S$ as Hilbert spaces that this quotient action
is the one on the
one-photon space of QED in the Coulomb gauge. The theorem then follows from the
well-known properties
of the latter, or by explicit computation. \hfill $\blacksquare$

 Hence in $x$-space $J^{-1}(0)$ consists of those fields which satisfy the
Lorentz
condition $\partial^{\mu}A_{\mu}(x)=0$ (weak derivatives), which is the precise
sense in which this
formalism implements Gauss' law at this stage.

The symplectic form $\om^0$ on $S^0$ (determined by the \MW\ reduction
procedure) may be described
as in subsection 2.2.: we now define an inner product on $S^0$ by
\be
([B],[C])_{S^0}= (P_TB,P_TC)_S \ll{om4}
\ee
in terms of the usual inner product on $S=\lt$, and obtain
\be
\om^0([B],[C])=-2\, {\rm Im}\,
([B],[C])_{S^0}. \ll{om5}
\ee
 Consequently, the
quotient \rep\ $\Uph$ of $P$ is unitary.

As in the previous subsection, the Poisson algebra $\cal A$ of gauge-invariant
smooth functions on $S$
(which is a subalgebra of the classical field algebra $\cal F$ of all smooth
functions on $S$) is
represented on the reduced space $S^0$, and the quotient of $\cal A$ by the
kernel  of this \rep\ is
the algebra of observables of classical electromagnetism.

Finally, let us sketch the entirely analogous development for helicity $\pm 2$,
that is, linearized
Einstein gravity. Here the starting point is the covariant \rep\
$U^{0,+,(1,1)}$ of $P$, realized on
the space of symmetric tensor fields $h_{\mu\nu}(x)$ on Minkowski space, which
satisfy the wave
equation $\square h_{\mu\nu}=0$, with Fourier coefficients in
$L^2(\R^3,d'p)\ot\C^{10}$.
The gauge group $G$ now consists of those non-constant weak solutions
$\xi_{\mu}(x)$ of the wave
equation for which the quantity $(\dl\xi)_{\mu\nu}=\partial_{\mu}\xi_{\nu}+
\partial_{\nu}\xi_{\mu}-g_{\mu\nu}\partial\cdot\xi$ lies in $S$.
$G$ acts on $S$ by $h\raw h+\dl\xi$. With the symplectic structure on $S$ given
by
$$\om(h,k)={\rm Im}\, \int d'p\, (h_{\mu\nu}(\bop)\ovl{k}^{\mu\nu}(\bop)-\half
h^{\mu}_{\mu}(\bop)\ovl{k}^{\nu}_{\nu}(\bop)),$$
 the group action is strongly Hamiltonian, with moment map given
by $J_{\xi}(h)=\om(\dl\xi,h)$. Hence $J^{-1}(0)=\{h\in
S|\partial^{\mu}h_{\mu\nu}=0\}$, and
$S^0=J^{-1}(0)/G$, which is the space of physical degrees of freedom of
gravitational waves, carries
the \rep\ $U^{0,+,2}\oplus U^{0,+,-2}$ of $P$.
 \subsection{Canonical and covariant Poisson actions of the Poincar\'{e} group}
This subsection provides a conceptual link between the preceding theory, which
describes classical
massless
field theory and quantum one-particle theory, and the theory of massless
classical particles.
It
may be skipped without losing the main thread of the paper.

We show that the
Hilbert space theory of the \rep s of $P$ has an analogue at the level of
symplectic manifolds, which
play the r\^{o}le of  phase spaces of classical relativistic particles. The
main idea here is that the
classical analogue of a unitary \rep\ of $P$ is a strongly Hamiltonian action
of $P$ on a symplectic
manifold \ci{AM,GS,Woo} (also cf.\ the Introduction above); the requirement of
irreducibility of a
Hilbert space \rep\ is then replaced by the transitivity of the action. A
well-known theorem of
Kostant and Souriau (see \ci{AM,GS,Woo}) then asserts that an irreducible
symplectic manifold for $P$
must be a co-adjoint orbit of $P$, or a covering space thereof; we will refer
to such manifolds
(along with the action of $P$ on them) as canonical realizations.

{\em 1. Canonical realizations.}
In the semidirect product $P=L \ltimes \rf$, $L$ will again stand for
$L_+^{\uparrow}$ in what
follows;  the inclusion of time inversion and parity is complicated also in
classical mechanics,
where the former is represented as an anti-canonical transformation. See
\ci{Woo} for a careful
treatment of the full Poincar\'{e} group in this context.

We wish to derive an expression analogous to (\ref{2.2}) for the co-adjoint
action of $P$ on an
orbit in the dual $\bop^*$ of its Lie algebra $\bop$.   The co-adjoint action
$\al_{\rm co}$ of
$(\Lm,a)\in P$ on $(M,p)\in \bop^*=\bol^*\oplus \rf$ is then given by
\ci{GS,MRW}
\be
\pco(\Lm,a)(M,p)=(\Lm M+ \vth_a(\Lm p),\Lm p) ,\ll{co}
\ee
where $\Lm M$ stands for the co-adjoint action of $\Lm$ on $M$, and
$(\Lm p)_{\mu}=\Lm_{\mu}^{\mbox{ }\rh}p_{\rh}$. The element $\vth_a(p)\in
\bol^*$
is defined by $\langle \vth_a(p),X\ra=\la p,Xa\ra$.
It follows that the co-adjoint orbit $\O^P_{(M,p)}$ through $(M,p)$ is
characterized by the $L$-orbit
of $p$ in $\rf$ and the $L_p$-orbit in $\bol^*_p$ through $\pi_p(M)$, where
$\pi_p:\bol^*\raw
\bol^*_p$ is given by restriction to $\bol_p$. In what follows we assume that
$p^2\geq 0$ and
$p^0>0$, so that we may take  $p$ to be our favourite point $\pt\in O_m\equiv
O_m^+$ (the theory is
identical for $O_m^-$). As before, we then have $\Lt=SO(3)$ for $m>0$ and
$\Lt=E(2)$ for $m=0$.
(The connection between these orbits and the canonical \rep s of $P$ can be
made explicit by either
geometric quantization \ci{Raw,Rob} or by the reduction-induction theory of
\ci{NPL}).

Various descriptions of the orbits under this action exist \ci{GS,Woo,NPL}, but
here we find it
convenient to exploit the fact   (proved in \ci{MRW} for general semidirect
products)  that the orbit
$\O^P_{(\til{M},\til{p})}$ through $(\til{M},\til{p})$
is realized as the symplectic leaf of the Poisson manifold $(T^*L)/\Lt$ (where
$\Lt$  acts on the
cotangent bundle $T^*P$, equipped with its canonical Poisson structure, by
pulling back its
right-action on $L$)  which corresponds to the co-adjoint orbit
$\O^{\Lt}_{\pi_{\pt}(\til{M})}$ in
$\bol^*_{\til{p}}$. To avoid cumbersome notations, we will simply refer to the
latter orbit in
$\bolt^*$ as $\O^{\Lt}_{\sg}$, where $\sg$ is some label characterizing the
co-adjoint orbits of
$\Lt$.  The leaf $L^{m,\sg}$ in question can be written as
\be
L^{m,\sg}=L\times_{\Lt} \pi_{\pt}^{-1}(\O^{\Lt}_{\sg}); \ll{leaf}
\ee
here (evidently) $\pi_{\pt}^{-1}(\O^{\Lt}_{\sg})\subset \bol^*$, and $\Lt$ acts
on $T^*P\simeq
L\times \bol^*$ by $R_h(\Lm,\th)=(\Lm h^{-1},\pco(h)\th)$.
The points of $L^{m,\sg}$ are then equivalence casses $[\Lm,\th]$, where
$(\Lm,\th)\sim
R_h(\Lm,\th)$ for all $h\in\Lt$. Note that $\Lt$ indeed maps
$\pi_{\pt}^{-1}(\O^{\Lt}_{\sg})$ into
itself under the co-adjoint action.

 Assuming that $\pi_{\pt}(\til{M})\in
\O^{\Lt}_{\sg}$, we will denote $\O^P_{(\til{M},\til{p})}$ by $\O^{m,\sg}$.
 The symplectomorphism $\rh: \O^{m,\sg} \raw L^{m,\sg}$ is given by
\be
\rh(M,\Lm\pt)=[\Lm,\Lm^{-1}M]; \ll{sym}
\ee
note that both sides are independent of the choice of $\Lm$ in the class $\{\Lm
h\}_{h\in\Lt}$.
The fact that $\rh$ is indeed a symplectomorphism relative to the canonical
(`Lie-Poisson-Kirillov-Kostant-Souriau-Arnold') symplectic form on $\O^{m,\sg}$
\ci{GS,AM} and the one
on $L^{m,\sg}$ inherited from $T^*L$ is proved in \ci{MRW}.

It follows from
(\ref{co}) and (\ref{sym}) that the action  $\pco^{m,\sg}=\rh\circ\pco\circ
\rh^{-1}$ of $P$ on
$L^{m,\sg}$ is given by
\be
\pco^{m,\sg}(\Lm,a)[\til{\Lm},\th]=[\Lm\til{\Lm},\th+\th_a(\Lm\til{\Lm})],
\ll{act}
\ee
where $\th_a(\Lm)=\vth_{\Lm^{-1}a}(\pt)$, i.e.,
$\la\th_a(\Lm),X\ra=\la\pt,X\Lm^{-1}a\ra$.
Note that $\pi_{\pt}(\th_a(\Lm))=0$, so that $\th+\th_a(\Lm\til{\Lm})$ in
(\ref{act}) indeed lies in
$\pi_{\pt}^{-1}(\O^{\Lt}_{\sg})$ if $\th$ does. Choosing a section
$\et:L/\Lt\simeq O_m\raw L$, as in
subsection  2.1, leads to a local trivialization of $L^{m,\sg}$, regarded as a
bundle over $O_m$.
In this trivialization $P$ acts by
\be
(\pco^{m,\sg})^{\et}(\Lm,a)(p,\th)=(\Lm p, \pco(\et(\Lm p)^{-1}\Lm
\et(p))\th+\th_a(\et(\Lm p))).
\ll{trivact}
\ee
We will not use this expression in what follows, but it is the classical
analogue of (\ref{2.2}).
  The
`cocycle' $\et(\Lm p)^{-1}\Lm \et(p)$ takes values in $\Lt$, so that the action
is well-defined.

Since $P$ acts transitively on $L^{m,\sg}$, its symplectic form $\om^{m,\sg}$
is fully determined by
specifying its value on the vector fields $\til{X}$ tangent to the flows
$\pco^{m,\sg}(\exp\, tX)$,
$X\in \bop$. Using  (\ref{sym}) and the canonical symplectic structure on
$\O^{m,\sg}$ we obtain
\be
\la \om^{m,\sg}|\til{X},\til{Y}\ra([\Lm,\th])=\la \Lm\th+\Lm\pt|[X,Y]\ra
.\ll{kk}
\ee

 An important special case is $\sg=\{0\}$, in which case the well-known
diffeomorphism $L\times_{\Lt}\bolt^{\perp}\simeq T^*(L/\Lt)$ (where
$\bolt^{\perp}\subset\bol^*$ is
the annihilator of $\bolt$) \ci{KN}, plus a short computation showing that the
canonical symplectic
form on $ T^*(L/\Lt)$ pulls back to  the form (\ref{kk}) on $L^{m,0}$  under
this diffeomorphism,
implies that $L^{m,0}\simeq T^*O_m$ as symplectic manifolds.

Another pleasant special case pertains when $m>0$ (and $\sg$ arbitrary; this
label now specifies the
radius of a two-sphere $S^2$ in ${\bf so(3)}^*=\R^3$). For in this case $\bol$
has the reductive
decomposition $\bol={\bf so(3)}\oplus \m$, where ${\bf so(3)}$ is generated by
the rotation
generators $\hat{J}_i$, and $\m$ is the span of the boost generators
$\hat{K}_i$ ($i=1,2,3$).
Reductive here means that $[{\bf so(3)},\m]\subseteq \m$,  and this property
implies that
the principal bundle $(L,L/\Lt,\Lt)$ has an $L$-invariant connection, viz.\ the
one defined by the
reductive decomposition \ci{KN}. This is important, for any connection on this
bundle
may be used to define a projection $pr:L^{m,\sg}\raw T^*O_m$ \ci{GS}, and
therefore a splitting of
$L^{m,\sg}$ into orbital and spin  degrees of freedom. The fact that the
connection is $L$-invariant
then means that this splitting is Poincar\'{e} covariant. In this case, one
even has the global form
$L^{m,\sg}=T^*O_m\times S^2_{\sg}$, which is well-known, except that in the
literature \ci{Woo,GS}
the first factor $T^*O_m$ is written as $T^*\rt$, where $\rt$ is $x$-space;
that is, the r\^{o}le of base
space and fibre is reversed.
 We leave it to the reader to apply the beautiful formalism of \ci{GS,MMR} to
the case at hand, and
obtain the Poisson bracket on $L^{m,\sg}$ as the sum of the canonical ones on
$T^*O_m$ and $S^2$, and
an extra term depending on the curvature of the connection.

For our main concern is the massless case, which is dramatically different from
the massive one.
In this case there exists no reductive decomposition $\bol={\bf e(2)}\oplus\m$,
and the orbital
degree of freedom does not factorize in a meaningful (i.e., covariant) manner.
 Recalling that ${\bf e(2)}$ is generated by
$\hat{T}_1=\hat{K}_1-\hat{J}_2,\hat{T}_2=\hat{K}_2+\hat{J}_1,\hat{J}_3$, the
most useful decomposition
is to
 take $\m$ to be
  the linear span of $\hat{K}_1+\hat{J}_2,\hat{K}_2-\hat{J}_1,\hat{K}_3$
(which, incidentally, in
contrast to the massive case, closes under Lie bracket, and  generates the Lie
algebra of Bianchi
type $V$; in this classification the Lie algebra of $E(2)$
   is of
type $VII_0$, and that of $SO(3)$ of type $IX$).

In any case, the co-adjoint orbits of $E(2)$ which are of interest to us are
the points
$\O_{\pm 1}$, whose value on $\hat{T}_1$ and  $\hat{T}_2$ vanishes, whereas the
value on $\hat{J}_3$
is $\pm 1$. These orbits  are the classical analogues of the helicity $\pm 1$
\rep s $U_{\pm 1}$ of
$E(2)$, but note that the co-adjoint action of $E(2)$ on these orbits is
(evidently) trivial, whereas
the action of $U_{\pm 1}$ on $\C$ is not. Nonetheless, the corresponding orbits
$\O^{0,\pm
1}\simeq L^{0,\pm 1}$ of $P$  are not equivalent to $L^{0,0}$ as realizations
of $P$. We shall refer
to the values $\pm 1$ as helicity.

{\em 2. Covariant  realizations.}
The covariant  symplectic realizations of $P$ \ci[sect.\ I.20]{GS} are
initially defined on symplectic
spaces of the type $T^*\rf\times \O^L_{a,s}$, where $\O^L_{a,s}$ is a
co-adjoint orbit in $\bol^*$ of the
Lorentz group $L$, equipped with the canonical   symplectic structure.
We follow \ci{GS} in identifying both the Lie
algebra $\bol$ of $L$ and its dual $\bol^*$ with the space of antisymmetric
$4\times 4$ matrices
$M_{\mu\nu}$; the pairing between $\bol^*$ and $\bol$ is given by contraction
using the Minkowski
metric $g$. Then $(\Lm M)_{\mu\nu}=\Lm_{\mu}^{\mbox{ }\rh}\Lm_{\nu}^{\mbox{
}\sg}M_{\rh\sg}$.

Then such a co-adjoint orbit
is characterized by the conditions \ci{GS,Woo} $\half M_{\mu\nu}M^{\mu\nu}=s^2$
and
$|\ep^{\mu\nu\rh\sg}M_{\mu\nu}M_{\rh\sg}|=24a$.
  In physically
relevant cases
$a=0$, and later we will only examine the orbit where
$s=1$, which leads to spin or helicity 1. In terms of the variables
$K_i=M_{0i}$ and $J_i=\half
\ep_{ijk}M_{jk}$ this means that $\bJ\cdot\bK=0$ and $|\bJ|^2-|\bK|^2=1$.

  We take $\om=dx^{\mu}\wedge dp_{\mu}$ as the symplectic form on $T^*\rf$
(where $p_{\mu}$ are the fiber co-ordinates on $T^*\rf$);
   the total
space has the product symplectic structure, and  $P$ acts on it by the product
$\al^{a,s}$ of the
pull-back of its defining action on $\rf$ and the co-adjoint action of $L$ on
$\O^L_{a,s}$. This
action is obviously strongly Hamiltonian, but it is `reducible' because it is
not transitive.

The first step in the `reduction' of $T^*\rf\times \O^L$,
is to impose the constraint $p^2=m^2$ (and $p_0>0$), and quotient the
constraint surface
$\rf\times O_{\til{p}}$ by the null
foliation of the induced presymplectic form. This is the same as performing a
\MW\ reduction w.r.t.\
the action of $\Bbb R$ on $T^*\rf$, given by $(x^{\mu},p_{\mu})\raw
(x^{\mu}+p^{\mu}\ta,p_{\mu})$.
The reduced space may be identified with $T^*O_m$, which we realize as
$L\times_{\Lt}\bolt^{\perp}$,
as before:  a point $[x,\Lm\pt]$ in the reduced space (brackets refering to the
$\R$-foliation)
may be identified with $[\Lm,\th_x(\Lm)]\in L\times_{\Lt}\bolt^{\perp}$
(brackets refering to
$\Lt$-equivalence classes). On account of the property $\th_x(\Lm
h)=h^{-1}\th_x(\Lm)$ for all
$h\in\Lt$, this identification is well-defined.

The $P$-action $\al^{m,a,s}$ on $S^{m,a,s}=T^*O_m\times \O^L_{a,s}$ is given by
the reduction of
$\al^{a,s}$; we   obtain
\be
\al^{m,a,s}(\Lm,a)([\til{\Lm},\th],M)=([\Lm\til{\Lm},\th+\th_a(\Lm\til{\Lm})],\Lm M).\ll{covact}
\ee
Using the previous observation that $T^*O_m\simeq L^{m,0}\simeq \O^{m,0}$, we
find that  the
symplectic structure $\om^{m,a,s}$ on $S^{m,a,s}$ has a particularly simple
form when evaluated on
the vector fields generating the $P$-action. For $X,Y\in \bop$ one has
\be
\la \om^{m,a,s}|\til{X},\til{Y}\ra([\Lm,\th],M)=\la \Lm\th+M+\Lm\pt|[X,Y]\ra
.\ll{symcov}
\ee
Since $P$ does not act transitively, this does not determine $\om^{m,a,s}$, but
the above expression
is sufficient for our purposes.

We now specialize to the case $a=0, s=1$, and attempt to make $P$ act
transitively by imposing a
further constraint (this was first done in \ci[I.20]{GS} in the massive case,
using a slightly
different formalism).  The appropriate constraint is $p^{\mu}M_{\mu\nu}=0$,
which in the present
setting comes from the function $\Phi:S^{m,0,1}\raw \rf$, defined by
$\Phi([\Lm,\th],M)=M\Lm\pt$
(which is indeed independent of the chosen representative of the class
$[\Lm,\th]$).
The constraint is then $\Phi=0$; as we shall see, this stands for two
independent conditions only.
The solution set $C_m$ is given by
\be
C_m=\{([\Lm,\th],M)\in S^{m,0,1}|M\in \Lm(\bolt\cap \O_{0,1}^L)\}, \ll{sol}
\ee
which is well-defined, for $\Lt$ maps $\bolt\cap \O_{0,1}^L$ into itself, so
that changing $\Lm$ by
$\Lm h$, $h\in \Lt$, alters nothing.

For $m>0$, the set $\bolt\cap \O_{0,1}^L$ consists of those $(\bJ,\bK)\in\bol$
for which $\bK=0$ and
$|\bJ|=1$. Consequently, $C_m$ is symplectic (in the traditional language of
constraint theory
\ci{Dir,Sun}, the constraints are second-class), as is evident from the fact
that we can find a
symplectomorphism $\phv:C_m\raw L^{m,1}$, given by
\be
\phv([\Lm,\th],M)=[\Lm,\th+\Lm^{-1}M]. \ll{scl}
\ee
 That this is a diffeomorphism is immediate from the definition
$L^{m,1}=L\times_{\Lt}\pi_{\pt}^{-1}(S^2_1)$ and the above description of
$\bolt\cap \O_{0,1}^L$;
that $\phv$ is symplectic is equally immediate from (\ref{kk}) and
(\ref{symcov}).
Moreover, the appropriate $P$-actions are correclty intertwined by $\phv$, in
the sense that
$\phv\circ\al^{m,0,1} =\pco^{m,1}\circ \phv$.

For $m=0$  the set  $\bolt\cap \O_{0,1}^L$ is given by
$\{(\bJ,\bK)\in\bol|K_1=-J_2,K_2=J_1,J_3=\pm 1,K_3=0\}$, so that $C_0$ is the
union of two components
$C_0^{\pm}$, the $\pm$ corresponding to the sign of $J_3$. This time, however,
the analogue of the
map (\ref{scl}) is not a diffeomorphism. For $\bolt\cap \O_{0,1}^L$ is
span$(\hat{T}_1,\hat{T}_2)\pm \hat{J}_3$, whereas $\bolt^{\perp}={\rm
span}(\hat{T}_1,\hat{T}_2,K_3)$. This reflects the phenomenon that
$\bolt^{\perp}\oplus \bolt\neq
\bol$ for $m=0$ (i.e., $\bolt={\bf e(2)}$), whereas equality does hold for
$m>0$ (where $\bolt={\bf
so(3)}$). Thus, for example, the points
$([1,(T_1+\lm_1,T_2+\lm_2,K_3)],(T_1'-\lm_1,T_2'-\lm_2,J_3=\pm 1))$ have the
same image under $\phv$
 for any $\lm_i\in\R$.

More systematically, the pull-back of the symplectic form $\om^{0,0,1}$ to
$C_0^{\pm}$ is not
symplectic; $C_0^{\pm}$ are co-isotropically embedded in $S^{
0,0,1}$, and the null directions of its presymplectic form may be found by
computing the
transformations generated by the constraint $\Phi$, or by a local computation
using (\ref{symcov})
(it follows from the explicit structure of $C_0^{\pm}$ that $P$ acts
transitively on each
component).
 Looking at $\Lm=1$ to start with,
  the four conditions $\Phi=0$ amount to  $\Phi_1=K_1+J_2=0,\,
\Phi_2=K_2-J_1=0$, and
$\Phi_0=\Phi_3=K_3=0$. Now recall that the orbit $\O^L_{0,1}$ was specified by
the conditions
$\bJ\cdot\bK=0$ and $|\bJ|^2-|\bK|^2=1$, which on imposition of the first two
constraints imply the
third one, which is therefore superfluous. Thus we need to identify two
independent constraints in
$\Phi$, and this may be done by choosing a global piecewisely smooth section
$\et:O_0\raw L$.
 The two independent constraints are then given by $\Phi^{\et}_1$ and
$\Phi^{\et}_2$,
where $\Phi^{\et}([\Lm,\th],M)=\et(\Lm\pt)^{-1}\Phi([\Lm,\th],M)$.
(If the constraints are written as $p^{\mu}M_{\mu\nu}=0$, the need to specify a
section $\et$
comes from the fact that one has to specify an isomorphism between
$p^{\perp}/\C p$ and $\R^2$ at
each $p$.)

The independent constraints generate an action $\bt$ of $\R^2$ on $S^{0,0,1}$,
which on $C_0$ is given
by \be
\bt_{\lm} ([\Lm,\th],M)=([\Lm,\th+\Lm^{-1}\et(\Lm\pt)\lm],M-\et(\Lm\pt)\lm).
\ll{rtact}
\ee
Here $\lm\in\R^2$ is identified with $\lm_1\hat{T}_1+\lm_2\hat{T}_2\in {\bf
e(2)}\subset \bol$,
on which $L$ acts by the co-adjoint \rep, as before. We stress that the
linearized form
(\ref{rtact}) only holds on $C_0$; the transformation of points off $C_0$ has a
  more complicated
form, which is  not needed in what follows. Since $\Lm^{-1}\et(\Lm\pt)\in \Lt$,
and $\Lt$ maps
$\bolt^{\perp}$ into itself under the co-adjoint action, and since $M-\lm\in
\bolt\cap \O_{0,1}^L$ if
$M\in  \bolt\cap \O_{0,1}^L$, the transformation (\ref{rtact}) indeed maps
$C_0$ into itself,
proving that  $C_0$ is co-isotropically embedded in $S^{0,0,1}$, as claimed.

Although the action of $\R^2$ is not defined canonically (in the sense that the
section $\et$ was
needed to specify it), its orbits are independent of $\et$, and define a
foliation of $C_0$.
A local computation verifies that this is precisely the null foliation ${\cal
F}_0$ with respect to
the induced presymplectic form.  The leaf space ${\cal L}=C_0/ {\cal F}_0$ of
this foliation
has two components ${\cal L}^{\pm}=C_0^{\pm}/{\cal F}_0$. We now recall the
definition of the
co-adjoint orbits $\O^{0, \pm 1}\simeq L^{0,\pm
1}=L\times_{E(2)}\pi_{\pt}^{-1}(\O_{\pm 1})$ of the
Poincar\'{e} group, which correspond to helicity $\pm 1$. If one inspects
(\ref{scl}),
in which $\phv$ is  now regarded as as a map from $C_0^{\pm}$ to $ L^{0,\pm
1}$, one immediately sees
that $\phv\circ \bt_{\lm}=\phv$ (with $\bt_{\lm}$ defined by (\ref{rtact})),
and that two points only
have the same image under $\phv$ if they are thus related. Hence $\phv$
quotients to map $\til{\phv}:
 C_0^{\pm}/{\cal F}_0\raw L^{0,\pm 1}$.
It follows from (\ref{kk}) and (\ref{symcov}) that $\til{\phv}$ is a
symplectomorphism.

We now show that $\til{\phv}$ intertwines the action of $P$.
Firstly, it is immediate from   (\ref{covact}) and (\ref{sol}) that $P$ maps
$C$ into itself.
Furthermore, we see
from (\ref{covact}) and (\ref{rtact}) that
\be
\pi^{0,0,1}(\Lm,a)\circ
\bt_{\lm}([\til{\Lm},\th],M)=\bt_{h(\Lm,\til{\Lm})\lm}\pi^{0, 0,1}(\Lm,a)
([\til{\Lm},\th],M), \ll{intertw}
\ee
where $h(\Lm,\til{\Lm})=\et(\Lm\til{\Lm}\pt)^{-1}\Lm \et (\til{\Lm}\pt)$, which
lies in $E(2)$.
This relation implies that $P$ acts on $C$ in a leaf-preserving way, so that
its action
quotients to the leaf space; we call this action $\til{\al}^{0,0,1} $. Finally,
(\ref{covact}) and
(\ref{scl}) lead to the conclusion that
 $\pco^{0,1}\circ \til{\phv}=\til{\phv}\circ \til{\al}^{0,0,1} $.

To sum up, we have proved
\begin{theorem}
Let $O_0=L/E(2)=\{p\in\rf|p^2=0,p^0>0\}$, and let $\O^L_{0,1}$ the co-adjoint
orbit of $L$ in
$\bol\simeq \bol^*$ characterized by the conditions  $\bJ\cdot\bK=0$ and
$|\bJ|^2-|\bK|^2=1$.
Then the symplectic reduction of $T^*O_0\times \O^L_{0,1}$ (equipped with the
product of the
respective  canonical symplectic structures) with respect to the constraint
$p^{\mu}M_{\mu\nu}=0$
($p\in O_0,M\in\O^L_{0,1}$) is symplectomorphic to $\O^{0,1}\cup \O^{0,-1}$,
where $\O^{0,\pm 1}$ is
the co-adjoint orbit of $P$ (equipped with the canonical symplectic structure)
characterized by
$m=0$ and helicity $\pm 1$. Moreover, the natural action of $P$ on
$T^*O_0\times \O^L_{0,1}$
survives this reduction, and is intertwined with the co-adjoint action on
$\O^{0,1}\cup \O^{0,-1}$
by the appropriate symplectomorphism.
\end{theorem}
The above procedure suggests that the reduced space is a \MW\ quotient with
respect to an action of
$\R^2$. Unfortunately, the action in question was not canonically defined, and
in the way it was
defined is only piecewisely smooth. A more intrinsic way to proceed would be to
consider the vector
bundle $L\times_{E(2)} \R^2$, where $\R^2$ is regarded as the abelian subgroup
of $E(2)$, as before;
the action of $E(2)$ on $\R^2$ is given
  by the restriction of the co-adjoint \rep\ (that is, $\R^2$ itself acts
trivially and $SO(2)$ acts by rotations).  This vector bundle is to be regarded
as a Lie groupoid,
which coincides with the associated Lie algebroid, and it acts on $T^*O_0\times
\O^L_{0,1}$ in the
sense of groupoid actions \ci{Mac}. One may then apply the generalized
symplectic reduction procedure
with respect to symplectic groupoid actions \ci{Xu,NPL} to reobtain the above
results.

 In any case, the parallel between the way one passes
from covariant to canonical massless Hilbert space \rep s, and symplectic
realizations,
respectively, is  striking. In both cases the essential point is the need to
quotient out
the directions (in Hilbert or symplectic space) in which the undesired abelian
subgroup of $E(2)$
acts nontrivially.
\section{Rieffel induction for the frozen photon field}
In this section we freeze the electromagnetic potential $\am$ at the value
$\til{\am}=\am(\pt)$, and
quantize the model with 4 degrees of freedom of  subsection 2.2.
\subsection{Tuning up}
For simplicity, we write $\ps_{\mu}$ for $\til{\am}$. Recalling that $S=\cf$
and that the symplectic
form on $S$ is given by (\ref{minisym}), we define the `field algebra' $\F$ of
the model to be
$\F=\W(S,\om)$, the Weyl algebra of canonical commutation relations (CCR)
defined by $S$ and $\om$.
This is generated by unitary elements $W(\ps)$ satisfying
 $W(\ps)W(\phv)=\exp(-\half i \om(\ps,\phv))W(\ps+\phv)$
(cf.\ \ci{BR2} or, in the present context, \ci{CGH1}).
For heuristic considerations it is useful to introduce the (unbounded)  bosonic
creation-  and
annihilation operators with commutation relations
$[a_{\mu},a_{\nu}^*]=-g_{\mu\nu}$, in terms of
which formally  $W(\ps)=\exp(a(\ps)-a(\ps)^*)$ (where
$a(\ps)=a_{\mu}\ovl{\ps}^{\mu}$, hence
$a(\ps)^*=a_{\mu}^*\ps^{\mu}$). In view of this, we will henceforth write $\ps$
with an upper index.

The gauge transformation generated by $\lm\in G=\C$ is given  by the inner
automorphism
$\al_{\lm}[A]=W(-\lm\pt)AW(-\lm\pt)^*$:
for if we take $A=W(\ps)$ we find from the CCR that
$\al_{\lm}[\exp(a(\ps)-a(\ps)^*)]=\exp((a+\lm\pt)(\ps)
-(a+\lm\pt)(\ps)^*)$.
The algebra of weak observables $\ovl{\A}$ is by definition the gauge-invariant
subalgebra $\F^G$ of
$\F$; equivalently, it is the commutant $\W(T,\om)'$ in $\F$ of the abelian
algebra
$\W(T,\om)=C^*(G_d)$ (here the commutant is defined as the collection of all
elements of $\F$ which
commute with all elements in the given subalgebra $\W(T,\om)$).
Cf.\ subsect.\ 2.2 for the definition of $N$ and $T$; $G_d$
stands for the group $G$ equipped with the discrete topology.
 The following equality \ci{GH3}  explicitly identifies the weak algebra of
observables.
\be
\W(T,\om)'=\W(N,\om). \ll{arakieq}
\ee
This result is valid for arbitrary $CCR$ algebras (of the minimal type defined
in \ci{Bel}). The
essential point is that $N$ is the symplectic orthoplement $T^{\perp}$ of $T$.
Here  the symplectic orthoplement $M^{\perp}$  of a subspace $M\subset S$ is
defined as
$$M^{\perp}=\{\phv\in S|\om(\phv,\ps)=0\:\forall
\ps\in M\}.$$

 Eq.\ (\ref{arakieq}) is Theorem 4.2 in  \ci{GH3}, whose proof unfortunately
contains a gap.
The proof given below was arrived at after a correspondence with H. Grundling
(intending to correct
the proof in \ci{GH3}).
\\
{\em Proof of Eq.\ (\ref{arakieq}), after H. Grundling}.
Things are made transparent if we look at the CCR algebra $\W(S,\om)$ as the
$C^*$-completion of the
twisted convolution algebra  $C^{\om}_c(S_d)$, where $S_d$ stands for $S$
equipped with the discrete
topology; the  twisted convolution product is defined on $C^{\om}_c(S_d)$
(which is $C_c(S_d)$ as a
space; the upper index $\om$ denotes the twist in the convolution) by
$(fg)(\ps)=\sum_{\phi\in S}
f(\phi)g(\ps-\phi)\exp(-\half i \om(\phi,\psi))$, and extended to
$C^*_{\om}(S)\equiv \W(S,\om)$ by
continuity, cf., e.g., \ci{Bel}. We then have the inequalities
\be
\n f\n_{\infty}\; \leq\;  \n f\n_2 \; \leq \; \n f\n   \ll{beeq}
\ee
where the first norm is the supremum one, the second norm is  in $L^2(S_d)$
 (with respect to the
counting measure on $S_d$), and the third is in
$C^*_{\om}(S_d)\equiv\W(S,\om)$. The first inequality is obvious (given the
discreteness of the
underlying measure space), the second  follows from the existence of the
tracial state $\om_0$,
defined
 by continuous extension of $\om_0(f)=f(0)$ \ci{Bel}; indeed, $\n
f\n_2^2=\om_0(f^* f)$.
It follows that $C^*_{\om}(S_d)$ as a Banach space (with its $C^*$-norm) is
continuously embedded in
$C_0(S_d)$ (with sup norm), for any element of the former is the limit of a
Cauchy sequence in
$C_c(S_d)$; by (\ref{beeq}) this sequence must also converge in the sup norm,
so that its limit must
lie in $C_0(S_d)$.

Now take an arbitrary $f\in C^*_{\om}(S_d)$, and a Cauchy sequence $f_n$ in
$C_c(S_d)$
converging to $f$ in $ C^*_{\om}(S_d)$.
Then (recalling \ci{Bel} that $W(\phv)=\dl_{\phv}$, the delta function with
support at $\phv$)
it follows from the CCR that
$[f_n,W(\phv)]$ is the function $f_n^{(\phv)}:\psi\raw 2if_n(\psi-\phv)\sin
(-\half\om(\ps,\phv))$.
Now $\lim_n f_n^{(\phv)}$ exists in $ C^*_{\om}(S_d)$, hence in $C_0(S_d)$.
The function $\psi\raw \sin (-\half\om(\ps,\phv))$ lies in $C_b(S_d)$, which is
the multiplier
algebra of $C_0(S_d)$. Hence $f_n^{(\phv)}\raw f^{(\phv)}$ (defined like
$f_n^{(\phv)}$, with $f_n$
replaced by $f$) in $C_0(S_d)$. By uniqueness of the limit, we infer
$f_n^{(\phv)}\raw f^{(\phv)}$
in $ C^*_{\om}(S_d)$. We conclude that $[f,W(\phv)]=f^{(\phv)}$.

Now $f$ is in $\W(T,\om)'$ iff  $[f,W(\phv)]$ vanishes for all $\phv\in T$. The
preceding paragraph
then yields $\n f^{(\phv)}\n =0$, whereupon  (\ref{beeq}) implies that
$f^{(\phv)}$ identically
vanishes for such $\phv$. Therefore (evaluating $f^{(\phv)}$ at
$\ps=\ps'+\phv$, and using
$\om(\phv,\phv)=0$), $f$ must vanish whenever its argument does not lie in
$T^{\perp}=N$, and
(\ref{arakieq}) follows. \hfill $\blacksquare$

(Note that Araki \ci[Theorem 1(5)]{Ara} gave an arduous proof of the
corresponding von Neumann
algebra result $\pi(\W(N,\om))''=\pi(\W(T,\om))'$ for
any regular \rep\ $\pi$ (where the commutants now have their usual meaning).
Since $\pi(\W(T,\om)')=\pi(\W(T,\om))'\cap \pi(\W(S,\om))$, whose bicommutant
is  $\pi(\W(T,\om))'$,
this result follows immediately from (\ref{arakieq}). Moreover, the regularity
assumption may
evidently be dropped.)

Thus $\ovl{\A}=\W(N,\om)$, and note that $\W(T,\om)$ is the centre of
$\ovl{\A}$.
 The notation $\A$ is reserved for the
dense subalgebra of $\ovl{\A}$ consisting of finite linear combinations of
elements $W(\ps)$, $\ps\in
N$ (hence  $\A=\W_0(N)$).

Refering to the Introduction or to \ci{NPL} for motivation, we now wish to
quantize the reduction
procedure of subsection 2.2 using Rieffel induction. As we have explained, in
the special case at
hand we first need a \rep\ $\pi(\F)$ on a Hilbert space $\H$, which carries a
unitary \rep\ $U$ of
$G$ as well. The latter point is actually taken care of by putting
$U(\lm)=\pi(W(-\lm\pt))$.

We essentially follow \ci{CGH1} in taking the so-called Fermi \rep\ $\pi_F$ of
$\F$. This is defined on
$\H=\exp(\cf)$ (the symmetric Hilbert space, or bosonic Fock space, over $\cf$
\ci{Gui,BR2}); on this
space the  usual creation-  and annihilation operators are defined, which we
denote by $\hat{a},\,
\hat{a}^*$;  their commutation relations are
$[\hat{a}_{\mu},\hat{a}_{\nu}^*]=\dl_{\mu\nu}$. The
heuristic idea of the Fermi \rep\ is to represent $a_i$ by $\hat{a}_i$, but
$a_0$ by $\hat{a}_0^*$,
so that the distinction between $-g_{\mu\nu}$ and $\dl_{\mu\nu}$ in the CCR is
taken care of.

A rigorous definition is arrived at by starting from the dense subset $E$ of
$\H$, which consists of
finite linear combinations of so-called exponential vectors \ci{Gui}. These are
defined on any symmetric Hilbert space $\exp({\cal K})$, and given for
arbitrary $\ps\in{\cal K}$ by
\be
e^{\ps}=\Om+\ps+\frac{1}{\sqrt{2!}}\ps\ot\ps+\ldots
+\frac{1}{\sqrt{n!}}\ps\ot\ldots \ps+\ldots ,
\ll{exp} \ee
where $\Om=\exp(0)$ is the vacuum vector in $\exp({\cal K})$, and $\ps\ot\ldots
\ps$ stands for the
tensor product of $n$ copies of $\ps$. We write this as
$\exp(\ps)=\sum_{n=0}^{\infty} (n!)^{-\half}
\ps^{\ot n}$. Note the square root (perhaps rendering the notation $\exp(\ps)$
somewhat
inappropriate); it guarantees that $(\exp(\ps),\exp(\phv))=\exp(\ps,\phv)$.

Back to ${\cal K}=\cf$, we define $\pi_F$ by linear extension of
\be
\pi_F(W(\ps))e^{\phv}=e^{-\half(\ps,\ps)-(\phv,\ps')}e^{\phv+\ps'}, \ll{pif}
\ee
where $\ps'\mbox{}^{\mu}=(\ovl{\ps^0},-\ps^i)$. To arrive at this expression,
simply use the heuristic idea
explained above, the BCH-formula, and the relations \ci{Gui}
$\exp(\hat{a}(\ps))\exp(\phv)=\exp(\phv,\ps)\exp(\phv)$,
$\exp(\hat{a}(\ps)^*)\exp(\phv)=\exp(\phv+\ps)$ (note that $E$ is
in the domain of $a(\ps)$ and $a(\ps)^*$, as well as of their exponentials
\ci{BR2}). It follows that
$\pi_F$ is indeed a \rep, which may then be extended to all of $\H$ by
continuity.
 (Our discussion
differs
  from \ci{CGH1} in that we have not altered the conventional complex structure
on $\cf$, but rather
interchanged the role of $a_0$ and $a_0^*$,
and also in our explicit construction of $\pi_F$ using exponential vectors.)

 We note that $\pi_F(\A)$ (which may be identified with $\A$, defined in the
second par.\ of this
subsection) leaves $E$ stable, hence we may attempt to perform Rieffel
induction on $L=E$. The \rep\
of $G$ on $\H$ is given by \be
U(\lm)=\pi_F(W(-\lm\pt)), \ll{grep}
\ee
and $\pi_F(\A)=\pi_F(\F)^G$, i.e., the set of elements of $\F$ which commute
with all $U(\lm)$.

If we would really need to apply Rieffel induction in its original formulation
\ci{Rie,FD}, we would
face a pesky dilemma at this point: if we choose $\B=C(G)\cap L^1(G)$, then the
rigging map
  $\la\ps,\phv\rab:\lm\raw (U(\lm)\phv,\ps)$ indeed takes values in $\B$, but
unfortunately
$\B$ (acting on $\H$ through $U$, cf.\ the Introduction) does not leave $E$
stable. On the other
hand, if we change the topology on $G$ and try $\B= L^1(G_d)$, then $\B\subset
\A$, so that
$E$ is stable under its action, but this time the rigging map does not take
values in $\B$.
Fortunately, all we need is formula (\ref{1.2}) for the rigged inner product
(in which $G$ has its
usual, Euclidean topology); this formula follows from the first choice of $\B$
mentioned, and the difficulty with the stability of $L$ can simply be ignored.
\subsection{The rigged inner product}
Thus we regard $L=E$ as a left-$\A$ module via $\pi_F$, and induce from the
trivial \rep\ of $G$.
The first step in the induction procedure is the introduction of the
sesquilinear form $\rip$ on
$L$. From (\ref{1.2}), this is given by
\be
(\Psi,\Phi)_0=\int_{\C}\frac{d\lm\, d\ovl{\lm}}{2\pi i}(U(\lm)\Psi,\Phi).
\ll{ipmm}
\ee
Using (\ref{grep}) and (\ref{pif}), we find that for $\Psi,\Phi\in E$ this
integral indeed converges,
and on elementary vectors the result is
\be
(e^{\ps},e^{\phv})_0=\exp(-\ps^0\ps^3-\ovl{\phv}^0\ovl{\phv}^3+\ps^1\ovl{\phv}^1+\ps^2\ovl{\phv}^2).
\ll{ripexp}
\ee
For later use, we note that we may rewrite this expression as
\be
(e^{\ps},e^{\phv})_0=(e^{\ps^{\perp}},\Om)_0\,(\Om,e^{\phv^{\perp}})_0\,
e^{(\ps^T,\phv^T)},
\ll{rewr}
\ee
where $\ps^{\perp}=(\ps^0,0,0,\ps^3)$ and $\ps^T=(0,\ps^1,\ps^2,0)$; in fact
$(\exp(\ps^{\perp}),\Om)_0 =(\exp(\ps),\Om)_0$.

Since $G$ is amenable, it follows from \ci[Proposition
2]{NPL} that property 1 (\ref{P1}) is satisfied, whereas property 3 (\ref{P3})
follows from
\ci[Proposition 2]{NPL}; note that these results were proved for $\B=C_c(G)$
rather than $C(G)\cap
L^1(G)$, but the proofs are not changed by this modification; moreover, the
only input in the proofs
is the fact that the rigging map takes values in $\B$, so that the stability of
$L$ under $\B$
(which, as we have discussed, is not satisfied in our application) was not
used. Property 2
(\ref{P2}) is immediately verified, since each $A\in\A$ commutes with $U(\lm)$.
Later on we will, in fact,
deduce (\ref{P1}) and (\ref{P3}) directly.

We now examine some properties of $\rip$. Regarded as a quadratic form on $\H$
with domain $E$, it
is positive (hence symmetric), but not closable. For if it were, there would
exist a positive
self-adjoint operator $A$ such that $(\Psi,\Phi)_0=(A\Psi,\Phi)$. From
(\ref{ripexp}) with $\phv=0$
we then find that $A\Om=\sum_{n=0}^{\infty} (n!)^{-1}\sqrt{(2n)!}\,
\boe_0\ot_s\ldots
\boe_0\ot_s\boe_3\ldots\ot_s\boe_3$ (the $n$'th term contains $n$ copies of
$\boe_0$ and $n$ copies
of $\boe_3$, and $\ot_s$ stands for the symmetrized tensor product, normalized
such that
$\ps\ot_s\ldots \ps=\ps\ot\ldots\ps$). But $(A\Om,A\Om)=\sum_n 1=\infty$, so
$A$ does not exist.

However, the domain of $\rip$ can be extended to the linear hull of $E\cup F$,
where $F$ is the dense
subspace in $\H$ consisting of finite linear combinations of vectors with a
finite number of
particles, i.e., of the type $\ps_1\ot_s\ldots \ps_n$, $n<\infty$. The
extension is obtained from
(\ref{rewr}) and the formula \be
\ps_1\ot_s\ldots\ps_n=\frac{1}{\sqrt{n!}}\frac{d}{dt_1}\ldots\frac{d}{dt_n}e^{\sum_{I=1}^n
t_I\ps_I}\mbox{}_{|t_1=\ldots t_n=0}. \ll{handy}
\ee
The most convenient expression for it is obtained by remarking that any vector
in $F$ is a finite
linear combination of vectors of the type $\ps_1^{\perp}\ot_s\ldots
\ps_l^{\perp}\ot_s
\ps_{l+1}^T\ot_s\ldots \ps_n^T$, and for those we obtain
\bea \lefteqn{
(\ps_1^{\perp}\ot_s\ldots \ps_l^{\perp}\ot_s
\ps_{l+1}^T\ot_s\ldots \ps_n^T,\phv_1^{\perp}\ot_s\ldots \phv_{l'}^{\perp}\ot_s
\phv_{l'+1}^T\ot_s\ldots \phv_{n'}^T)_0=} \nn \\
& &
(\ps_1^{\perp}\ot_s\ldots \ps_l^{\perp},\Om)_0\,
(\Om,\phv_1^{\perp}\ot_s\ldots \phv_{l'}^{\perp})_0\,
(\ps_{l+1}^T\ot_s\ldots \ps_n^T,\phv_{l'+1}^T\ot_s\ldots \phv_{n'}^T). \nn \\ &
& \ll{ipE}
\eea
This expression vanishes if the number of transverse components does not match
(i.e., if $n-l\neq
n'-l'$), or if $l$ or $l'$ are odd. For $l=2m$ even, one has
\be
(\ps_1^{\perp}\ot_s\ldots\ps^{\perp}_{2m},\Om)_0=\frac{(-1)^m}{m!\sqrt{(2m)!}}\sum_{P\in
S_{2m}}\ps_{P(1)}^0\ldots\ps_{P(2m)}^3, \ll{pso}
\ee
where $m$ $\ps$'s on the r.h.s.\ carry the upper index $0$, and $m$  $\ps$'s
carry $3$ upstairs.
The sum is over all permutations of $\{1,\ldots,2m\}$.
Equation  (\ref{pso})  is still valid if each $\ps_i^{\perp}$ is replaced by
$\ps_i$.
 The expression (\ref{ipE}) may alternatively
be derived directly from (\ref{ipmm}), with  $\Psi$ and $\Phi$ chosen in $F$.
A general formula for the rigged inner product on $F$ is given in \ci{UAW}.

The anatomy of $\rip$ is then clear from the observation that
$\ps_1^{\perp}\ot_s\ldots \ps_l^{\perp}\ot_s
\ps_{l+1}^T\ot_s\ldots \ps_n^T$ equals
$(\ps_1^{\perp}\ot_s\ldots \ps_l^{\perp},\Om)_0\,
\ps_{l+1}^T\ot_s\ldots \ps_n^T$ plus a vector in the null space ${\cal N}$ of
$\rip$. Hence the
non-transverse components merely provide a numerical factor to the transverse
ones, up to null
vectors.

Finally, we remark that $\rip$ does not preserve the adjoint: for example,
the property  $(\pi_F(W(\ps))\Psi,\Phi)_0=(\Psi, (\pi_F(W(-\ps))\Phi)_0$ only
holds if $\ps\in N$.
\subsection{The induced \rep}
Let us now return to Rieffel induction on the domain $L=E$.
Recalling the discussion of the classical situation in subsection 2.2, our aim
is to show that the
induced space $\H^0$ (which  is the completion of $L/{\cal N}$) is naturally
isomorphic
to the bosonic Fock space $\exp(S^0)$. Remember that $S^0=J^{-1}(0)/G=N/T$; in
this case this
is isomorphic to $\C^2$ as a Hilbert space, the inner product being given by
(cf.\ (\ref{om2}))
 \be
([\ps^T],[\phv^T])_{S^0}=(\ps^T,\phv^T)_{S}. \ll{ipnt}
\ee
Here $S=\cf$,
and $[\ps^T]$ is the image of   $\ps$  under the double projection map, which
first projects
$\ps$ onto its component in $N$, and then onto $N/T$.
 We now attempt to define a map $V:L\raw \exp(S^0)$ by linear extension of
\be
Ve^{\ps}=(e^{\ps},\Om)_0\, e^{[\ps^T]}=e^{-\ps^0\ps^3}\, e^{[\ps^T]}, \ll{V}
\ee
 There is a subtle   point here:
we could have  identified $\C^2$ with span$\{\boe_1,\boe_2\}$ rather than with
$N/T$,
and define  $V$ by writing $\ps^T$ for $[\ps^T]$ in (\ref{V}). In view of
(\ref{ipnt})
this would
  equally well satisfy the crucial relation  (\ref{uni}) below; the properties
of the induced
\rep\ $\pi^0(\A)$ would then compel us to re-identify $\C^2$ as $S^0=N/T$
anyway, as will become clear
shortly.

Another subtlety is that the basis $\{\exp{\ps}\}$ is overcomplete, so that it
is not obvious that
the linear extension of (\ref{V}) is well-defined. To see that it is, consider
the map
$\hat{V}:F\raw \exp(S^0)$, defined by
\be
\hat{V}\, \ps_1^{\perp}\ot_s\ldots \ps_l^{\perp}\ot_s
\ps_{l+1}^T\ot_s\ldots \ps_n^T=\left(\frac{(n-l)!l!}{n!}\right)^{\half} (
\ps_1^{\perp}\ot_s\ldots \ps_l^{\perp},\Om)_0  \,[\ps_{l+1}^T]\ot_s\ldots
[\ps_n^T],
\ll{hatV}
\ee
extended to $F$ by linearity; this extension is manifestly well-defined. This
map $\hat{V}$ is
unbounded, but it can be extended to exponential vectors;  a short calculation
shows
that     this extension  coincides with $V$; hence $V$ may  consistently be
extended to $E$ by linearity, and may be thought of as being defined on the
linear hull of $E\cup F$.
The point is that it follows from (\ref{V}) and (\ref{rewr})  that \be
(V\Psi,V\Phi)=(\Psi,\Phi)_0, \ll{uni}
\ee
where the inner product on the l.h.s.\ is the one in $\exp(S^0)$.
This firstly verifies  that $\rip$ is postive semi-definite on $L$, and
secondly that its null
space $\cal N$ coincides with ${\rm ker}\, V$. Therefore, $V$ may be extended
to $\H^0=\ovl{L/{\cal
N}}$, and  provides a unitary isomorphism $\til{V}$ between the induced space
$\H^0$ and
$\exp(S^0)$.

The induced \rep\ $\pi^0$ is therefore characterized by
\be
\pi^0(A)\circ V=V\circ \pi_F(A) \ll{inter}
\ee
 for all $A\in\A$. Since $\pi_F(\A) $ is a \rep\ and $\til{V}$ is unitary, this
also shows,
incidentally, that $\pi^0$ is a \rep\ of $\A$, so that the bound (\ref{P3}) is
satisfied.
The intertwining relation (\ref{inter}) allows us to  describe $\pi^0$
explicitly.
For it follows from (\ref{pif}), (\ref{V})
 that $\pi^0(W(\ta))=1$ for all
$\ta\in T$;
 another way to
say this is that for all $\Psi\in L$,  one has $\pi_F(W(\ta))\Psi=\Psi+ \nu$,
where $\nu\in {\cal
N}$. That is, gauge transformations leave vectors in $L$ invariant up to null
vectors.
 Subsequently, the property $W(\ps+\ta)=W(\ps)W(\ta)$ for all $\ps\in
N$ and $\ta\in T$ implies that $\pi^0(W(\ps+\ta))=\pi^0(W(\ps))$.
 Moreover, we
see that $\pi^0(W(\ps^T))$ coincides with the Fock representative of
$W([\ps^T])$ on $\exp(S^0)$.

We now define   the CCR algebra $\W(S^0,\om^0)$ over $S^0$ with the usual
symplectic
structure $\om^0$ (cf.\ (\ref{om3})),    and define
$\pi_{\Phi}$ to be the Fock \rep\ of $\W(S^0,\om^0)$ on $\exp(S^0)$
\ci{Gui,BR2}. Then
(recalling that $N=J^{-1}(0)$ and $S^0=J^{-1}(0)/G$) the
properties of $\pi^0$ mentioned imply that
\be
 \pi^0(\W(J^{-1}(0),\om))\simeq \pi_{\Phi}(\W(S^0,\om^0)), \ll{ident}
\ee
where we have extended the induced \rep\ $\pi^0$ from $\A$ to its completion
$\W(N,\om)$, which is
allowed because the bound (\ref{P3}) is satisfied \ci{Rie}. In particular,
since
$\pi_{\Phi}$ is faithful we find that $ \W(J^{-1}(0),\om)/{\rm ker}\,
\pi^0\simeq \W(S^0,\om^0)$,
where ${\rm ker}\, \pi^0$ is generated by elements in $\W(N,\om)$ of the type
$W(\ta)-1$, $\ta\in T$.
The algebra $\W(S^0,\om^0)$ is the algebra of observables of the model, and we
see that it is
obtained from the `field algebra' $\F=\W(S,\om)$ in two steps: first the
gauge-invariant subalgebra
$\F^G=\W(N,\om)$ is selected, and then the remaining gauge transformations are
eliminated by
building a quotient.

Finally, consider the action of the group  $E(2)$. As described in subsection
2.2, $E(2)$ acts on
$S= \cf$; this action is symplectic, and therefore leads to an automorphic
action of $E(2)$ on the
field algebra $\F$, defined by $\al_x[W(\ps)]=W(x\ps)$ (also cf.\ \ci{CGH1}).
 Since $N$ is invariant under $E(2)$, this automorphism group may be restricted
to $
\W(N,\om)$, and also to the  dense subalgebra $\A$ of the latter.

This automorphism group happens to be unitarily implementable in $\pi_F(\F)$,
but since   in the full
electromagnetic theory the corresponding property (for the Lorentz group) fails
to hold, we will
not exploit it. Rather, the relevant fact is that the generating functional of
the induced \rep\
$\pi^0(\A)$ is invariant: for
 \be
(\pi^0(W(\ps))\Om^0,\Om^0)=(\pi_F((W(\ps))\Om,\Om)_0=e^{\half(\ps,\ps)_M},
\ll{eti}
 \ee
where $\Om^0=V\Om$ is the vacuum vector in $\exp(S^0)$, and
$(\ps,\ps)_M=\ps^{\mu}\ovl{\ps}_{\mu}$, as before.  This follows from
(\ref{pif}) and (\ref{ripexp})
if one uses $\ps^0=\ps^3$  for all $\ps\in N$. Therefore, we can define a
unitary \rep\ $U^0$ of
$E(2)$ in the usual way \ci{BR1}, viz. $U^0$ is given by extending the equation
$U^0(x)\pi^0(W(\ps))\Om^0=\pi^0(W(x\ps))\Om^0$ for all $\ps\in N$  by linearity
and continuity.
It may be checked \ci{UAW} that this \rep\ coincides with the second
quantization of the \rep\
$U_1\oplus U_{-1}$, cf.\ subsection 2.2.

We invite the reader to compare our discussion with the treatment of this model
in \ci{CGH1}.
There the \rep\ $\pi^0(\W(N,\om))$ is constructed in a completely different
way, which
is arguably a bit {\em ad hoc} compared with our induction procedure. Moreover,
one finds in
\ci{CGH1} a whole family $\pi^{\xi}$ of \rep s of $\W(N,\om)$, where $\xi\in
\C$.
In our formalism, \rep s unitarily equivalent to these may be obtained by
treating $\xi$ as a character of the gauge group $G=\C$, and
inducing not from the trivial \rep\ of $G$, but from the one defined by $\xi$,
cf. \ \ci{UAW} for details.
\section{Rieffel induction in electromagnetism}\eo
\subsection{Preamble}
We now treat the entire field $\am(x)$ in the manner of the preceding section.
The setting is almost
exactly as in section 3.1, with the following changes (cf.\ subsection 2.3, or
\ci{CGH1}, in which
ref.\ $S$ stands for our $P_TS$). The phase space $S$ is now given by $\lt$,
with symplectic form
(\ref{2.7}). We continue to denote its elements by $\ps,\,\phv$. The closed
subspaces $N$ and $T$ of
$S$ are now defined by
\be
N=\{  A\in S|p_{\mu}\ps^{\mu}(\bop)=0 \, ({\rm a.e.})\}; \:\:\:\:
T=\{d\lm|\lm\in G\}, \ll{NT}
\ee
where the gauge group $G$ has been defined prior to (\ref{Sob}); recall our
symbolic notation
$d\lm$ for the element of $S$ defined by
$(d\lm)^{\mu}(\bop)=-ip^{\mu}\lm(\bop)$, $p^0=|\bop|$.
We still have $T^{\perp}=N=N^{\perp\perp}$.
Also, in the \MW\ reduction of $S$, $J^{-1}(0)$ coincides with $N$, and
$S^0=J^{-1}(0)/G=N/T$, cf.\
subsection 2.3.
   The
field algebra is $\F=\W(S,\om)$, and since   (\ref{arakieq}) holds without
modification, $\ovl{\A}=\W(N,\om)$ is again the commutant in $\F$ of
$\W(T,\om)=C^*(G_d)$, which in
turn is contained in $\ovl{\A}$ as its center. Gauge transformations in $\F$
are given by
automorphisms $\al_{\lm}[W(\ps)]=W(d\lm)W(\ps)W(d\lm)^*$, $\lm\in G$, so that
$\ovl{\A}=\F^G$.

A new feature, indicating that we are now in the setting of quantum field
theory, is that, as
explained in \ci{CGH2} (following the discussion in \ci{Kas} for scalar
fields), $\F$ has a subalgebra
with the structure of a local net of $C^*$-algebras in the sense of
Haag-Kastler (cf.\
\ci{Haa,Hor}).  Namely, the local algebra $\F(\O)$ ($\O\subset \rf$ open) is
obtained as the
$C^*$-closure of the subalgebra of $\F$ generated by those $W(\ps)$ for which
$\ps^{\mu}(\bop)$
possesses an extension $\til{\ps}^{\mu}(p)$ off the mass-shell $p^2=0$ whose
Fourier transform lies
in the Schwartz space ${\cal D}(\O)$.

The Fermi \rep\ of $\F$ on $\H=\exp(S)$ is again given by (\ref{pif}). (Note
that the
smeared operators $\hat{a}(\ps),\, \hat{a}(\ps)^*$ satisfy the commutation
relations
 $[\hat{a}(\ps),\hat{a}(\phv)^*]=(\phv,\ps)$, and act as indicated below
(\ref{pif})).
The \rep\ $U$ of the gauge group is now given by (cf.\ (\ref{grep}))
\be
U(\lm)=\pi_F(W(d\lm)). \ll{grep2}
\ee
\subsection{Functional integral \rep\ of $\rip$}
Our aim is to  construct a  \rep\ $\pi^0$ of $\A$ which is Rieffel-induced from
the trivial \rep\
$\pi_0$ of the gauge group $G$, and the Fermi \rep\ $\pi_F(\F\res \A)$ on
$\exp(S)$, restricted to the dense
subspace $L=E$ of exponential vectors (cf.\ (\ref{exp})).
 Thus we would now like to construct the rigged inner
product $\rip$ by a formula analogous to (\ref{ipmm}), replacing the integral
over the frozen gauge
group $\C$ by one over the infinite-dimensional  Hilbert Lie group $G$.  We
start by inspecting the
equation \bea
\lefteqn{
(U(\lm)e^{\ps}, e^{\phv})=\exp(-\n \lm\n_{\half}^2)\times } \nn \\
& & \exp\left( (\ps,\phv)+i[(p^0\ps^0,\ovl{\lm})+({\bf p}\cdot
\ps,\lm)+(\ovl{\lm},p^0\phv^0)+(\lm,\bop \cdot  \phv)]\right), \ll{matrel}
\eea
which follows from (\ref{grep2}) and (\ref{pif}); recall (\ref{Sob}).
 All the inner products occurring in the exponentials on the r.h.s.\ are in
$L^2(\R^3,d'p)$, and we use
the   notations $p^0\ps^0$ and  $\bop\cdot\ps$ to denote the functions defined
by
$(p^0\ps^0)(\bop)=|\bop|\ps^0(\bop)$,  and
$(\bop\cdot\ps)(\bop)=p^i\ps^i(\bop)$, respectively.
 These inner products are all finite, as can be seen from the precise
definition of the Hilbert
space $G$. In the following measure-theoretic considerations, $G$ is regarded
as a real Hilbert
space.

Inspired by (\ref{matrel}), we consider the standard Gaussian weak distribution
$\mu_*$ \ci{Sok}
(alternatively called a promeasure or cylindrical measure \ci{AMP}) on $G$.
This is defined by
first considering an arbitrary $n$-dimensional Hilbert subspace $\H_n\subset G$
($n<\infty$),
which has a measure $\mu_n$ defined on an arbitrary Borel set $A\in\H_n$  by
$\mu_n(A)=\pi^{-n/2}\int_A d^n \lm\,\exp(-\n \lm\n_{\half}^2)$.
The
orthogonal projection onto $\H_n$ is denoted by $P_n$. The
measure of the cylinder set  $P_n^{-1}(A)$ is then given by
 $\mu_*(P_n^{-1}(A))=\mu_n(A)$.
If we now choose an inductive family $\{\H_n\}_n$ of such $\H_n$ which
eventually exhausts $G$ (that
is, $\H_n\subset \H_{n+1}$, and the closure of $\cup_n\H_n$ is $G$; we say that
$G$ is the inductive
limit of the family; this does not mean that the topology on $G$ is the
corresponding inductive
limit topology), then the promeasure $\mu_*$ is eventually defined on all
cylinder sets in $G$ by the
equation above. Conversely, each finite-dimensional Hilbert subspace ${\cal K}$
(not necessarily
contained in the
inductive family) is then equipped with a Gaussian measure $\mu_{\cal K}$,
defined as
$\mu_{\cal K}(A)=\mu_*(P_{\cal K}^{-1}(A))$, where  $P_{\cal K}:G\raw {\cal K}$
is the
orthogonal projector onto ${\cal K}$.

The covariance of this
promeasure is the unit operator, which is not nuclear (given that $G$ is
infinite-dimensional), and
therefore $\mu_*$ cannot be extended to a Borel measure on $G$ \ci{Sok}.
However, `tame' functions
(also called cylinder functions) can be integrated with respect to $\mu_*$;
these are functions on
$G$ of the type $f(\lm)=f_{\cal K}(P_{\cal K}\lm)$, where $f_{\cal K}$ is a
Borel function on a finite-dimensional
Hilbert subspace ${\cal K}$ of $G$. Note that $f(P_{\cal K}\lm)=f(\lm)$.
  The `integral' $\int_G d\mu_*\, f$ is then by definition equal to   the
Lebesgue integral
$\int_{{\cal K}}d\mu_{\cal K} \, f_{\cal K}$.
 The usual theorems and rules of Lebesgue
integration theory often apply if only  such tame functions are involved
\ci{Sok}.

With this preparation, we can define and compute the rigged inner product on
$L=E$.
\begin{prop}
Choose an inductive family $\{\H_n\}_n$ of finite-dimensional Hilbert subspaces
of $G$, such that
$G$ is the inductive limit of this family. Let $\mu_*$ be the standard Gaussian
weak distribution
(promeasure) on $G$. Then
\be
(\Psi,\Phi)_0=\lim_n \int_{\H_n} \frac{d^n \lm}{\pi^{n/2}}\,
(U(\lm)\Psi,\Phi)\ll{limit}
\ee
exists for all $\Psi,\,\Phi\in E\subset \exp(S)$, and on elementary vectors
equals
\bea
\lefteqn{
(e^{\ps},e^{\phv})_0=e^{(\ps,\phv)}\int_G d\mu_*(\lm) } \nn \\
& &
\exp i[(p^0\ps^0,\ovl{\lm})+({\bf p}\cdot
\ps,\lm)+(\ovl{\lm},p^0\phv^0)+(\lm,\bop \cdot  \phv)].   \ll{proint}
\eea
\ll{existence}
\end{prop}
{\em Proof}.  Since $\Psi\in E$, we can write
$\Psi=\sum_{I=1}^nc_I\exp(\ps_I)$, $n<\infty$,
$c_I\in\C$, $\ps_I\in S$; an analogous expansion holds for $\Phi$. By
(\ref{matrel}),
$(U(\lm)\Psi,\Phi)= \exp(-\n \lm\n_{\half}^2)f(\lm)$, where $f$ is a tame
function on $G$,
for it  is a finite sum of terms of the form $g(\lm) =\exp((\mu_1,\lm)+\ldots
(\mu_l,\lm))$,
where the $\mu_I$ are of the form $(\pm i)\,p^0\ps_I^0$, $(\pm
i)\,\bop\cdot\ps_I$, etc.\ (recall
that $G$ is here regarded as a real Hilbert space, so   there is no overall
factor $i$).
Hence in the above definition of tame functions,  ${\cal K}$ is the Hilbert
space spanned
by $\mu_1,\ldots,\mu_l$.

By definition of $\mu_*$ and the projector $P_n$ (explained above), one has
$$\pi^{-n/2}\int_{\H_n}
d^n \lm\, (U(\lm)\Psi,\Phi) =\int_G d \mu_*(\lm)\,f(P_n\lm).$$
Now  $f(P_n\lm)$ is a finite sum of terms of the form
$g_n(\lm)=\exp((P_n\mu_1,\lm)+\ldots (P_n\mu_l,\lm))$, which are evidently
still tame, and
$\int_G d\mu_*\, g_n$ can be explicitly evaluated, cf.\ \ci{Sok}.  Since
$P_n\raw 1$
weakly by construction of the family $\{\H_n\}_n$,  it is immediately verified
that
$\lim_n\int_G d\mu_*\, g_n=\int_G d\mu_*\, g$. This implies the
proposition.\hfill $\blacksquare$

Note that the Lebesgue dominated
convergence theorem could not be used in this proof,
although $\lim_nf(P_n\lm) =f(\lm)$ pointwise, and each $f\circ P_n$ as well as
the limit
function $f$ are tame and  bounded by a $\mu_*$-integrable function. The reason
is that in order to apply this theorem, the limit function
 and (eventually) all functions in the sequence   should depend on a given set
of
vectors $\mu_I$, whereas in the above case these vectors depend on $n$.

 From (\ref{proint}) onwards, we can sail through. The Gaussian functional
integral can be computed
by naive methods, for the integrand is tame, with the result
 \be
(e^{\ps},e^{\phv})_0=e^{-(\ps^0,\ovl{\ps}^L)-(\ovl{\phv}^0,\phv^L)+(\ps^T,\phv^T)}, \ll{ripqem}
\ee
where $\ps^T=P_T\ps$, with $P_T:S\raw S$ the usual projector onto the
transverse (physical)
degrees of freedom (see (\ref{pt})),
and  $\ps^L=\bop\cdot \ps/|\bop|$.  The first two inner products in the
exponential at the r.h.s.\ are
in $L^2(\R^3,d'p)$, and the third one is in $\lt$. Comparing this result with
(\ref{ripexp}), we see
that $\ps^L$ and $\ps^T$ play the r\^{o}le of $\ps^3$ and $(\ps^1,\ps^2)$ in
the frozen   model,
respectively; the status of $\ps^0$ is unchanged. The symbol
$\ps^{\perp}\in\cf$ used in section 3
now stands for $(\ps^0,(p^ip^j/|\bop|^2)\ps^j)\in \lt$. The entire discussion
in subsections 3.2 and
3.3 may then be taken over with the obvious notational modifications.
\subsection{The induced \rep\ for electromagnetism}
Firstly, (\ref{ipE}) is still valid as it stands, but (\ref{pso}) is replaced
by
\be
(\ps_1^{\perp}\ot_s\ldots\ps^{\perp}_{2m},\Om)_0=\frac{(-1)^m}{m!\sqrt{(2m)!}}\sum_{P\in
S_{2m}}(\ps_{P(1)}^0,\ovl{\ps_{P(2)}}^L)\ldots(\ps_{P(2m-1)}^0,\ovl{\ps_{P(2m)}}^L),
\ll{pso2}
\ee
where the inner products are in $L^2(\R^3,d'p)$. With $N$ and $T$ as defined in
(\ref{NT}),
and $S^0=N/T$ equipped with the inner product (\ref{ipnt}), we
define the map $V:L=E\raw \exp(S^0)$ by the analogue of (\ref{V}):
\be
Ve^{\ps}=(e^{\ps},\Om)_0\, e^{[\ps^T]}=e^{-(\ps^0,\ovl{\ps}^L)}\, e^{[\ps^T]}.
\ll{V2}
\ee
(Alternatively, we could start from (\ref{hatV}).)
{}From (\ref{ripqem}) and  (\ref{ipnt})  we infer that (\ref{uni}) is satisfied
on $L$,
eventually leading to the
identification of the induced space $\H^0$ with $\exp(S^0)$.
The desired properties (\ref{P1})-(\ref{P3}) are verified in the same way as
before, i.e.,
(\ref{P1}) follows from (\ref{uni}), (\ref{P2}) is immediate from the
definition of $\A$, and
(\ref{P3}) follows from (\ref{inter}), or from (\ref{ident}). Alternatively,
they can be proved
directly from (\ref{limit}) and the results in \ci{NPL} for amenable locally
compact groups, for
each approximant in (\ref{limit}) is an integral over such a group (namely
$\R^n$), and the
inequalities (\ref{P1}) and (\ref{P3}) are obviously stable under taking the
limit in $n$.

As a first application of (\ref{V2}), we consider  the Hamiltonian $H_F$
on $\H$. It is defined as the operator which implements the time-evolution
 on $\F$ in the Fermi \rep\ $\pi_F$, in the sense that
$\pi_F(\al_t[A])=\exp(it H_F )\pi_F(A)\exp(-it H_F)$.
The
 time-evolution on $\F$ is given by the
one-parameter automorphism $\al_t$, defined  by $\al_t[W(\ps)]=W(
\exp(-itH^{(1)})\ps)$.
The operator $H^{(1)}$ appearing in this definition is the one-particle
Hamiltonian $H^{(1)}$ on $S$, given by
$(H^{(1)}\ps^{\mu})(\bop)=p^0\ps^{\mu}(\bop)$  on the
obvious domain $D(H^{(1)})$ (note that
 $H^{(1)}$ is already in diagonal form).
However, the peculiar factor $-g_{\mu\nu}$ (rather than the normal
$\dl_{\mu\nu}$)
in the CCR defining $\F$
leads to the perhaps somewhat surprising fact that $H_F=d\Gamma
(\til{H}^{(1)})$,
where $\Gamma$ is the second quantization operation \ci{RS2}, and
$\til{H}^{(1)}$ is
defined (and self-adjoint) on  $D(H^{(1)})\subset S$ by
$(\til{H}^{(1)}\ps^{\mu}(\bop)=(-p^0\ps^0
(\bop), p^0\ps^i(\bop))$. In other words,
$H_F$ is the closure of the operator (defined on those finite-particle states
in $F\subset \H$ whose
components lie in $D(H^{(1)})$)
 \be
H^{\rm core}_F=-g^{\mu\nu}\int d'p\, |\bop| \hat{a}_{\mu}(\bop)^*
\hat{a}_{\nu}(\bop), \ll{hamil}
\ee
where  $\hat{a}_{\mu}(\bop)^*$, $\hat{a}_{\nu}(\bop)$ are the usual quadratic
forms
associated to the creation- and annihilation operators on $\H$ (cf.\
\ci[X.7]{RS2}).

Clearly, the spectrum of $H_F$ is $\R$, but if we inspect (\ref{V2}) or
(\ref{hatV}), and recall
that the null space $\cal N$ of $\rip$ is ${\rm ker}\, V$, we infer that any
vector in $L$ equals a
vector with only transverse components plus a null vector; hence
\be
(H_F \Psi,\Psi)_0\,\geq 0\:\:\: \forall \Psi\in L, \ll{posit}
\ee
for we see from (\ref{hamil}) that $H_F$ has positive expectation value in
transverse states.
This `rigged' positivity property eventually implies that the Hamiltonian in
the induced \rep\
$\pi^0$  has positive spectrum.

We now return to the construction of the induced space.
In complete analogy with the frozen case,
the crucial fact is that  the gauge transformations
(\ref{grep2}) do not affect the rigged inner product $\rip$, in the sense that
$U(\lm)\Psi$ equals
$\Psi$ plus a null vector for all $\Psi\in L$.
We can reformulate this property in   a way that
clarifies the relation between our formalism and the `Fermi method' of
quantizing the
electromagnetic field (cf.\ci{CGH1}).
To do so, first note that $\pi_F(\F)$ is a regular \rep, so that the field
potentials $\am^F$ exist as
operator-valued distributions on $\H$. They are related to $\F$ by means of
\ci{CGH1}
$\pi_F(W(D*f))=\exp(iA^F(f))$, where $f\in {\cal S}(\rf)\ot\rf$, $A^F(f)=\int
d^4x\,
\am^F(x)f^{\mu}(x)$, and $D$ is the usual Pauli-Jordan commutator function.
Choosing
$f^{\mu}=\partial^{\mu}\lm$, the above property is equivalent to
\be
(\partial^{\mu}\am^F(x)\Psi,\Phi)_0=0 \ll{gb}
\ee
as a distribution, for all $\Psi,\Phi$ initially in $L=E$, and, as explained
earlier, extendable to
the linear hull of $E\cup F$.

It should be clear
that  the discussion following (\ref{inter}) can be copied here (with
$\ta=d\lm$); in particular,
$\pi^0(W(d\lm))=1$ for all $\lm\in G$.
With $\om^0$ now defined by (\ref{om5}), the identification
(\ref{ident}) is valid as it stands in the present setting. As we have seen in
subsection 2.3,
$S^0$ is the space of physical degrees of freedom of electromagnetism. This is
reinforced by showing
that it carries the correct \rep\ of the Poincar\'{e} group $P$, cf.\
(\ref{uph}).

Since $S$ carries a \rep\   $\Ugf$ of $P$ which leaves the symplectic form
$\om$ invariant (cf.\
subsection 2.3), we can define $P$ as an automorphism group on $\F$ by means of
$\al_x[W(\ps)]=W(x\ps)$, in somewhat symbolic but obvious  notation. This
automorphism group cannot
be unitarily implemented in $\pi_F$  \ci{CGH1} (though its subgroup
$SO(3)\ltimes \rf$ can).
However, (\ref{eti}) is valid also in the present case,  so that the state on
$\ovl{\A}=\W(N,\om)$ defined by $\Om^0$ and $\pi^0$ is Poincar\'{e}-invariant.
Hence we obtain a \rep\ $U^0$ of $P$ on $\H^0$ by the procedure explained after
(\ref{eti}), and it is
easily checked that $U^0$ is
 the second quantization of $\Uph$, the photon \rep\ which already emerged from
\MW\ reduction in
subsection 2.3.

We sum up in a theorem. The definitions of the symplectic space $(S,\om)$
(where $S=\lt$ as a Hilbert space), the gauge group $G$, and the \MW\ quotient
$S^0=J^{-1}(0)/G$
(with the symplectic form $\om^0$ on $S^0$)
are given in subsection 2.3; note that $S^0=N/T$, with $N=J^{-1}(0)$ and $T$
defined in (\ref{NT}).
\begin{theorem}
Take the field algebra $\F$  of quantum electromagnetism to  be the CCR-algebra
$\W(S,\om)$, on
which $G$ acts by inner automorphisms. Let the `algebra of weak observables'
$\F^G=\W(J^{-1}(0),\om)$
be its gauge-invariant subalgebra. With $\A$ the dense subalgebra of  $\F^G$
spanned by the Weyl
operators $W(\ps)$, $\ps\in
 J^{-1}(0)$, we obtain a left-$\A$ and right-$G$ module $L$ as follows:
$L$ is the subspace of $\exp(S)$ spanned by exponential vectors,
the left-action of $\A$ is given by the restriction of the Fermi \rep\
$\pi_F(\F)$ (cf.\ (\ref{pif}))
to $\A$, and the action $U$  of $G$ is given by $U(\lm)=\pi_F(W(d\lm))$ (see
(\ref{grep2})).

Using a generalized notion of Rieffel induction (in which the rigging map on
$L$ is replaced by the
direct definition of a rigged inner product (\ref{limit}) on it),  these data
can be used
to  construct a \rep\ $\pi^0(\F^G)$ induced from the
trivial \rep\ $\pi_0$ of $G$.
 Then $\pi^0(\W(S,\om)^G)\simeq \W(S^0,\om^0)$ as $C^*$-algebras, with respect
to the local
structure, and with respect to the automorphic action of the Poincar\'{e} group
$P$. Moreover,
  the induction procedure produces $\W(S^0,\om^0)$, which is the algebra of
observables of
quantum electromagnetism, in its
vacuum \rep\ on $\H^0=\exp(S^0)$, in which the action of $P$ can be unitarily
implemented. This
realizes the induced space $\H^0$ as a Fock space of physical photons.
\ll{endth} \end{theorem}
\section{Discussion}\eo
In this final section we briefly discuss certain isolated aspects  of our
formalism.
Our coverage is mostly incomplete and  tentative, and is intended to inspire
further work in this
direction.
\subsection{The use of quantum fields}
Our construction was based on  choosing  the domain $L$, on which
the rigged inner
product (\ref{limit}) is defined, to be
 $L=E\subset \H=\exp(S)$, the space of exponential vectors (coherent states for
physicists). This
  has the
disadvantage that $E\cap F=\C\,\Om$, where $F$ is the finite-particle subspace
of $\H$.
In other words, $E$
 does not contain  states  with a finite number of particles
(apart from the vacuum). Yet
 all computations of scattering amplitudes in physics start from states in $F$,
so if we wish
to entertain the hope that  our
method may be of some practical use, we should incorporate such states.
On the one hand, this is straightforward, for we have seen that the rigged
inner product (regarded
as a quadratic form on $\H$), albeit unclosable, can be extended to the linear
hull of $E\cup F$.
The problem lies in our use of the Weyl algebras $\F$ and  $\ovl{\A}=\F^G$. In
order  to
carry through the induction procedure,
it is necessary that $\F^G$ contains a dense subalgebra $\A$ which leaves $L$
stable. With $L=E$   we could take $\A$ as specified in the theorem above, but
for $L=F$ no such
subalgebra exists.

The simplest way out of this dilemma is to work with unbounded operators, using
the
Borchers-Uhlmann-Maurin formulation of algebraic quantum field theory (cf.\
\ci{Hor}, and refs.\
therein). The Borchers algebra  $\F_u$  appropriate to quantum electromagnetism
was defined and
analyzed in \ci{Bon}. Leaving a rigorous study to the future, one may expect
that an unbounded analogue of the Fermi \rep\ $\pi_F$ on $\exp(S)$ can be
defined, so that
$\pi_F(f_1\ot\ldots f_n)=A^F(f_1) \ldots A^F(f_n)$; we already encountered the
fields $\am^F$
in (\ref{gb}) and surrounding text. As already discussed in \ci{NPL}, one can
carry through Rieffel
induction for $Op^*$-algebras $\A_u$ of unbounded operators (cf.\ \ci{Hor}) as
long as $L$ is a common
invariant domain of $\A_u$, and the crucial property (\ref{P2}) is satisfied.
This is indeed the case if $L=F$, and $\A_u\subset \F_u$ (the subalgebra in
which test functions lie
in $N\cap {\cal S}(\rf)\ot\rf$) acts on $F$ through $\pi_F$.

The idea is then to compute physical amplitudes, which in theory can be
expressed as (squared) matrix
elements in $\H^0$, in $\H$, using the rigged inner product rather than the
original one on $F$.
The simplest object is the propagator $(T\am^F(x)\Om,A^F_{\nu}(y)\Om)_0$ (where
$T$ is the
time-ordering instruction), which is found to coincide with  the one in
Coulomb-gauge QED.
(Due to the fact that the $\am^F$ are not symmetric with respect to $\rip$, the
above expression
differs from the two-point function $(T\am^F(x)A^F_{\nu}(y)\Om,\Om)_0$.)
The main difficulty in setting up a perturbation theory for interacting models
along these lines
is that Wick's theorem does not hold, for the reason that $(A\Om,\Om)_0$ does
not necessarily
vanish for normal-ordered expressions $A$ in the creation- and annihilation
operators.
See \ci{UAW} for  details, and some further steps.
\subsection{Use of the temporal gauge}
Our choice of $(S,\om)$ for electromagnetism (cf.\ (\ref{2.7})) was motivated
by the connection
between canonical and covariant \rep s of the Poincar\'{e} group, and led to a
covariant formalism
  in all stages of the classical theory, and most stages (field algebra,
algebra of weak observables,
induced \rep, algebra of observables and its vacuum \rep) of the quantum
theory.
Alternatively, a non-covariant approach based on the partial gauge-fixing
$A_0=0$ is possible, and
suitable for certain applications (e.g, thermal field theory, topological
effects in nonabelian gauge
theories, Hamiltonian approach to anomalies, \ldots).

In that case, we take $S=T^*Q$ with its canonical symplectic structure $\om$,
where
$Q=L^2_{\R}(\R^3)\ot\R^3$ is the configuration space of the spatial field
$\bA$.
We use canonical co-ordinates $(A_i,E_j)$ on $S$.
The gauge group $G$ is the subspace of ${\cal S}'(\R^3)$ of distributions $\lm$
whose weak exterior
derivative $d\lm$ lies in $S$, modulo constants; in contrast to our previous
formulation, this is a
{\em real} Hilbert space without a complex structure. $G$ acts on $S$ by
$\bA\raw \bA+d\lm$; this
action is strongly Hamiltonian, with moment map $J:S\raw  \g^*\simeq G$  given
by $J(\bA,{\bf
E})=\Delta^{-1}\nabla\cdot {\bf E}$. Hence $J^{-1}(0)$ consists of those points
in $S$ where Gauss'
law $\nabla\cdot {\bf E}=0$ is satisfied, and the \MW\ reduced space
$S^0=J^{-1}(0)/G$ consists of
solutions of Gauss' law modulo gauge transformations (it is  symplectomorphic
to the space called $S^0$ in subsection 2.3.).

To quantize, we define the field algebra $\F=\W(S,\om)$, and represent it
in the Fock \rep\ $\pi_{\Phi}$ on the symmetric Hilbert space $\H=\exp(S)$
(identifying $S$ with $L^2(\R^3)\ot\C^3$).
The gauge group acts on $\H$ through the unitary \rep\ $U$, defined by
$U(\lm)=\pi_{\Phi}(W(0,d\lm))$.
We wish to produce an induced \rep\ of $\A$, which is the appropriate dense
subalgebra of
$\W(J^{-1}(0),\om)$.
Thus we choose $L=E\subset \H$, which is a left-$\A$ and right-$G$ module,
and  define a rigged inner product on $L$ much as in Proposition
\ref{existence}, cf.\
(\ref{limit}). The result is
\be
(e^{\ps},e^{\phv})_0=e^{(\ps^{L},\ovl{\ps}^{L})+(\ovl{\phv}^{L},\phv^{L})
+(\ps^T,\phv^T)}, \ll{ipa0}
\ee
where $\ps^T=P_T\ps$, with $P_T$ defined in (\ref{pt}), and
$\ps^{L}=P_{L}\ps$, with $P_{L}=1-P_T$. All inner products are in
$L^2(\R^3)\ot\C^3$.
Comparing this expression with  (\ref{ripqem}), we infer that the r\^{o}le of
the inner product
$(\ps^0,\ovl{\ps}^L)$ is now played by $(\ps^{L},\ovl{\ps}^{L})$. The anatomy
of $\rip$ is
still that all vectors in $L$ can be decomposed as the sum of a purely
transverse part and a null
vector.

The construction of the induced \rep\ in subsection 4.3 may then be adapted in
an obvious way.
The situation is actually simpler here, for the subtle distinction between
$P_TS$ and $N/T$ is
irrelevant now. Thus $S^0=P_TS$, so that the induced space $\H^0$ may be
identified with $\exp(P_T
S)$, and we may write $\ps^T$ rather than $[\ps^T]$ in the present version of
(\ref{V2}).
 One eventually recovers the results stated in Theorem \ref{endth} (replacing
the claims about the
Poincar\'{e} group by those about its subgroup $SO(3)\ltimes \rf$).
\subsection{Regularity}
An alternative $C^*$-algebraic procedure to handle constrained systems was
developed in \ci{GH1,GH2}.
Restricting our discussion to the application of this  so-called $T$-procedure
to systems with
first-class contraints, and  grossly simplifying the story, this
procedure is intended to resolve a spectral
difficulty faced by the traditional Dirac method \ci{Dir,Sun}. Namely, given a
\rep\ $\pi$ of the
algebra of operators $\F$ of the unconstrained system on a Hilbert space $\H$,
the prescription of
Dirac is to look for gauge-invariant vectors in $\H$, and obtain a \rep\ of the
algebra $\F^G$ of
gauge-invariant operators on the `physical' subspace $\H_{\rm phys}$ of $\H$
spanned by such vectors
by simply restricting  $\pi_F(\F \res\F^G)$ to $\H_{\rm phys}$. This
prescription only works if 0 is
in the discrete spectrum of all the constraints (with common corresponding
eigenvectors), which is
rarely the case.

For example, in the
context of electromagnetism, we take $\F$ and $\F^G$ as in the main text, and
$\pi=\pi_F$;
the physical states $\Psi$ should then satisfy $\pi_F(W(d\lm))\Psi=\Psi$ for
all $\lm\in G$.
This is impossible, and $\H_{\rm phys}$ as defined by Dirac is empty in this
case.
This example was generalized to a theorem in \ci{GH3}, which, applied to
electromagnetism,
states that  $\H_{\rm phys}$ is empty whenever $\pi(\F)$ is a regular \rep.

In  due fairness to physicists using the Dirac procedure (e.g., in quantum
gravity), it should be
pointed out that in practice the equation $C\Psi=0$ (where $C$ is a constraint)
is solved not as an
eigenvalue problem in Hilbert space, but as a partial differential equation for
which nothing is said
in advance about the space of solutions. The drawback of such a procedure is
obviously that an inner
product on the space of solutions has to be found from scratch.

In any case, the $T$-procedure starts from a \rep-independent definition
of physical states as those states $\om$ on $\F$ for which $\om(W(d\lm))=1$ for
all $\lm\in G$.
It follows that such `Dirac' states must be non-regular on $\W(S,\om)$, and
this has inspired an
approach to gauge theories  based on the use of \rep s in
which gauge-variant fields do not exist \ci{NT}. In the $T$-procedure, the
algebra of observables
$\A_{\rm obs}$,
which we obtained as $\pi^0(\F^G)$, is constructed in a \rep-independent way.
Subsequently, physical states which (after restriction and quotienting) are
regular
on $\A_{\rm obs}$ may be obtained from those Dirac states which are regular on
$\W(N,\om)$.

 We now briefly summarize how our approach manages to avoid the use of
non-regular states.
The traditional Dirac condition $\pi_F(W(d\lm))\Psi=\Psi$ is replaced by
$\pi_F(W(d\lm))\Psi=\Psi+\nu$, where $\nu$ has to lie in the null space $\cal
N$ of the rigged inner
product $\rip$. As we have seen, this condition is actually satisfied by {\em
all} vectors $\Psi$ in
the dense subspace $L\subset \H$ on which $\rip$ is defined.
The price one pays is that the functional $\ps$ on $\W(S,\om)$, defined by
$\ps(A)=(\pi_F(A)\Psi,\Psi)_0$ (where $\Psi\in L$), only defines a state on
$\W(N,\om)$. The reason
for this is that the positivity property $(\pi_F(A^*A)\Psi,\Psi)_0\geq 0$ fails
for general
$A\in\W(S,\om)$: the rigged inner product only preserves hermiticity for
$A\in\W(N,\om)$.

For the purpose of comparison with the $T$-procedure, our formalism may be seen
as a method of
constructing states on $\W(N,\om)$, hence eventually on the algebra of
observables, from states in a
regular \rep\ of the entire field algebra.
 The need to consider non-regular
states on $\F$ does not arise   at all. Clearly, the fact that our weakened
version of the Dirac condition is identically satisfied on $L$ means that the
non-physical state
vectors in the Dirac method (as well as in the $T$-procedure) play a different
r\^{o}le in our
approach. We do not need to exclude such state vectors by hand: their
non-invariance perishes
automatically when the induced space $\H^0$ is constructed.

These points are particularly clearly illustrated when $G$ is compact,
in which case $(\Psi,\Phi)_0=(P_0\Psi,P_0\Phi)$, where $P_0$ projects on the
subspace of
 $\H$ carrying the trivial
\rep\ of $G$
(cf.\ the Introduction). Then, according to both the $T$-procedure and the
Dirac method,
the  physical states in $\H$ are only those that lie in $P_0\H$. On the other
hand, all vectors in
$\H$  satisfy the condition $U(x)\Psi=\Psi+\nu$, $\nu\in {\cal N}$ for all
$x\in G$; here ${\cal
N}=(P_0\H)^{\perp}$.  The analogue of $\W(N,\om)$ is the algebra of
$G$-invariant bounded
operators on $\H$ (the `field' algebra now consisting of all bounded
operators).
The induced space is $\H^0=\H/{\cal N}=P_0\H$, and we see clearly how the
non-invariant part of each
non-Dirac state disappears.
 
 \end{document}